\documentclass[smallextended]{svjour3}       
\smartqed  
\usepackage{graphicx}
\usepackage[numbers]{natbib}
\usepackage{graphics}
\usepackage{amsmath}
\usepackage{amsfonts}
\usepackage{amssymb}
\usepackage{array}
\begin{document}

\title{Bifurcation analysis and multistability detection of two delay-coupled FHN neurons}


\author{Niloofar Farajzadeh Tehrani         \and
        MohammadReza Razvan 
}


\institute{F. Author \at
              Department of Mathematical Sciences, Sharif University of Technology, Tehran, Iran.  \\
              Tel.: +123-45-678910\\
              Fax: +123-45-678910\\
              \email{farajzadeh@mehr.sharif.ir}           
           \and
           S. Author \at
              Department of Mathematical Sciences, Sharif University of Technology, Tehran, Iran.}

\date{Received: date / Accepted: date}

\maketitle

\begin{abstract}
This paper presents an investigation of the dynamics of two coupled non-identical FitzHugh-Nagumo neurons with quadratic term and delayed synaptic connection. We consider coupling strength and time delay as bifurcation parameters, and try to classify all possible dynamics. Bifurcation diagrams are obtained numerically or analytically from the mathematical model, and the parameter regions of different behaviors are clarified.
The neural system exhibits a unique rest
point or three ones by employing the saddle-node bifurcation, when strong coupling is applied in the system. Also the trivial rest point shows transcritical bifurcation with one of the new rest points. The asymptotic stability and possible Hopf and Bautin bifurcations of the trivial rest point are studied by analyzing the corresponding characteristic equation. Fold cycle, torus, fold-torus, and big homoclinic bifurcations of limit cycles, together with $1:1$ and $1:4$ resonances are found. The delay-dependent stability regions are illustrated in the parameter plane, through which the double-Hopf, Hopf-transcritical, and double-zero bifurcation points can be obtained from the intersection of different bifurcation branches.
Various patterns of multistability have been observed, both for small and large values of delay.
 The system may exhibit one synchronous together with one or two anti-phase periodic activities, two synchronous and two anti-phase periodic solutions, one synchronous and one anti-phase periodic solution also one equilibrium, and one anti-phase periodic solution and non-trivial eguilibria, which occur due to Hopf, fold cycle and torus bifurcations. Also one synchronous periodic solution and one torus due to fold cycle, torus and Chenciner bifurcations can be seen. By increasing the range of parameter $\tau$ other branches of fold and torus bifurcations appear, which lead to creation of more stable periodic solutions.  
\end{abstract}
\section{Introduction}\label{intro}
Neurons and their interactions are generally assumed to be the
determinant of the brain performance. The simplest model to display features of neural interactions consists of two coupled neurons or neural systems. Starting from such simple and reduced networks, larger networks can be built and their features may be studied. In order to study complicated interaction between neurons in large neural
networks, the neurons are often put up into highly connected sub-networks or synchronized sub-ensembles. In this way, the model of two mutually coupled neurons may also apply as a framework of two coupled neural networks.
 Destexhe et al. \cite{destexhe1994model}, by investigating a model of spindle rhythmicity in the isolated thalamic reticular nucleus, showed that how more complex dynamics emanates in ring networks with fully mutual connectivity of nearest neurons, or in networks in which every neuron connects to all other nearby neurons.  Zhou et al. \cite{zhou2006hierarchical} modeled a neural network as a small sub-network of excitable elements, to study synchronization dynamics and the hierarchically  organization of excitable neurons in complex brain networks.\\
The FitzHugh-Nagumo (FHN) model with cubic non-linearity was derived as a simplified model of the famous Hodgkin-Huxley (HH) model \cite{hodgkin1952quantitative}. This model was suggested by FitzHugh (1961) \cite{fitzhugh1961impulses}, who called it “Bonhoeffer – van der Pol model”, and the equivalent circuit was constructed by Nagumo et al. (1962) \cite{nagumo1962active}. This model is a classic oscillator exhibiting variety of nonlinear phenomena in planar autonomous systems. The FHN-like systems are of fundamental importance for describing the qualitative nature of impulse propagation and neural activity. In fact, this model seems reach enough, and can capture neural excitability of original HH equation \cite{pouryahya2013nonlinear}. Moreover, it is versatile in the sense that as well as neurons it has also been used to model cells including those in the
 heart \cite{gray2002termination}, and other areas such as calcium oscillations \cite{timofeeva2003oscillations}, and even the pulsatile release of luteinizing hormone \cite{foweraker1995discrete}. In this way, Hoff et al. \cite{hoff2014numerical} investigated unidirectional and bidirectional electrical couplings of two identical FHN neurons. According to four different parameters, they studied bifurcations and synchronization numerically. In addition, different firing patterns such as chaotic firing is found in a pair of identical FHN elements with phase-repulsive coupling \cite{yanagita2005pair}. In modeling studies of trans-membrane potential oscillations, Bachelet et al. \cite{doss2003bursting} explained that bursting oscillations appear quite naturally for two coupled FHN systems.
Deng et al. investigated the response of three coupled FHN neurons, under high-frequency driving, to a sub-threshold low-frequency signal. They showed that the chemical synaptic coupling is more efficient than the well-known electrical coupling (gap junction), especially when the coupled neurons are near the canard regime \cite{deng2009effect}.
 The effect of different types of chemical synapses on vibrational resonance in coupled neurons is investigated by Chun-Hua et al. \cite{chun2012effects}. They studied three coupled FHN neurons, under high-frequency driving, to a sub-threshold low-frequency signal, and showed that with the chemical synapse ranging from slow to fast, the signal processing efficiency is at first increasing and then remain unchanged. The influence of time delay of different motifs on vibrational resonance is also analyzed in this paper. In \cite{FarajzadehTehrani201541} the authors studied the bifurcation structure of two coupled non-identical and symmetric FHN neurons. By considering $c$ and $\tau$ as bifurcation parameters many bifurcation branches are obtained. Multiple periodic solutions are shown in stated paper.\\
It is known that signal transmission in coupled neurons is not instantaneous in general \cite{swadlow2012axonal}. Hence a time delay (time latency) can occur in the coupling between neurons in some areas of brain or in a self-feedback loop. 
The finite  speed  of  signal transmission  over  a  distance  gives  rise  to  a  finite  conduction	 delay.
For example the speed of signal transmission through unmyelinated axonal fibers through the cortical network is in the order of 1 m/s, which leads to presence of time delays up to 80 ms for signal propagation \cite{swadlow2012axonal,kandel2000principles}.
On the other hand, to control neural disturbances, e.g. to suppress alpha rhythm \cite{hadamschek2006brain}, and undesired synchrony of firing neurons in Parkinson’s disease or epilepsy \cite{schiff1994controlling,rosenblum2004controlling,popovych2005effective}, time-delayed feedback mechanisms can be implemented intentionally. It is known that some neurological diseases are caused by synchronization of neurons. Therefore various delayed feedback loops have been proposed as effective and powerful therapy of neurological diseases which are due to synchrony \cite{rosenblum2004controlling,popovych2005effective,gassel2007time,gassel2008delay,dahlem2008failure,scholl2009time}.\\
According to above discussion the study of coupled FHN systems with delay has attracted many authors’ attention. Buric and Todorovic \cite{buric2003dynamics} investigated Hopf bifurcations, and also Bautin bifurcation \cite{buric2005bifurcations} of coupled FHN neurons with delayed coupling, in the case of small time lags. Buric et al.\cite{buric2005type}, by variation of the coupling strength and time delay observed different synchronization states in a delayed coupling of FHN neurons. They showed that the stability and the patterns of exactly synchronous oscillations depend on the type of excitability and type of coupling of neurons. For a system of two electrically delay-coupled neural system, Dahlem et al.\cite{dahlem2009dynamics} showed that, for sufficiently large delay and coupling strength bi-stability of a fixed point and limit cycle oscillations occur due to saddle-node bifurcation of limit cycles. The Fold-Hopf bifurcation is investigated in a coupled FHN neural system with delay by Zhen and Xu \cite{zhen2010fold}. Fan and Hong \cite{fan2010hopf} considered the stability and Hopf bifurcation of double delay coupled FHN neurons, see also \cite{xu2014dynamics}. The steady state bifurcations of two coupled FHN neurons due to coupling strength and small time delay is investigated in \cite{zhen2010simple} and \cite{rankovic2011bifurcations}. Zhen and Xu \cite{zhen2010bautin} studied Bautin bifurcation of completely synchronous three coupled FHN neurons with delay. Panchuk et al. \cite{panchuk2013synchronization} investigated effects of heterogeneous time delays for mutual and self-coupling of two coupled FHN system.\\
In addition to study of coupled FHN neurons, the dynamics of a network of coupled FHN neurons with delay is attractive for researchers. Buric and Todorovic \cite{buric2003dynamics} have analyzed examples of systems of identical FHN neurons arranged in linear or circular lattices, with uni- or bi-directional symmetrical coupling by some typical coupling models. Lin \cite{lin2014stability} studied a network of FHN neurons with delayed coupling and different synaptic strength of self-connection. They have described stability properties of
emerging periodic patterns. Perlikowski et al. \cite{perlikowski2010periodic} by investigating a network of coupled FHN neurons interacting via excitatory chemical synapses, have described stability properties of emerging periodic patterns. Kantner et al. \cite{kantner2015delay} by investigating two-dimensional lattices of coupled oscillators with in-homogeneous coupling delays, have shown that arbitrary stable spatio-temporal periodic patterns can be seen. They offer their model for the generation, storage, and information processing of visual patterns.
 The effects of heterogeneous coupling delays in complex networks of excitable FHN units is studied in \cite{cakan2014heterogeneous}. Two discrete delay times as well as uni- and bi-modal continuous distributions is investigated in referred paper.\\
 In the present paper, we investigate the effect of the coupling strength and time delay on the stability and bifurcations of the system of two synaptic coupled FHN neurons. The FHN model and parameters are chosen such that the neurons are a-symmetric, and without coupling both are at rest. We will show although the system (\ref{1}) and system (\ref{1}) in \cite{FarajzadehTehrani201541} are largely similar, their dynamical behavior can be different.
 While in many of the above mentioned papers, only local and
codimension-1 bifurcations of FHN neurons are investigated, the study of
global and codimension-2 bifurcations are crucial to classify different dynamical behaviors such as transition between synchronous and anti-phase solutions, and also simultaneous existence of stable periodic solutions. Therefore, applying bifurcation methods, we give a complete qualitative analysis of our model. Actually the study of bifurcation structure is a serious task, and without such analysis it is not easy to show under which circumstances there are stable limit cycles, and how many limit cycles the system have simultaneously. For example, in our work we could not detect some of stable periodic solutions just with simulations. Actually, by the guidance of bifurcation study, we first predicted their existence, then approved them with simulation.\\
To study the delay effects on neural system in details, in this paper,
we have analyzed Hopf, fold cycle, Bautin, double-Hopf, torus (Neimark–Sacker bifurcation in the Poincare map), fold-torus, big homoclinic, saddle-nod, transcritical, hopf-transcritical and double-zero bifurcations. Also various dynamical behaviors are classified in the neighborhood of a double-Hopf point. We will describe the creation of a stable torus due to Chenciner bifurcation.  We also show the existence of different modes of spiking as a result of fold cycle and torus bifurcations.
Since our neurons are assumed to be at rest without coupling, the periodic solutions are due to coupling of the neurons, or time delay. In this paper, we show the existence of both synchronized and anti-phase solutions, and also various patterns of multistability in different ranges of parameters.\\
The paper is organized as follows. A brief description of the FHN
neuron model is introduced in Section ~\ref{sec:fhnmodel}. Also the linear stability of the trivial equilibrium is investigated, and the critical values of $\tau$ for creation of Hopf bifurcations are driven. In Section ~\ref{sec:trivial equilibria}, with the aid of numerical simulations, bifurcation diagrams for trivial equilibrium are obtained. In Section ~\ref{sec:strong}, the strong coupling of neurons is considered and numerical simulations are carried out for determining the bifurcations of non-trivial equilibria which are appeared through saddle-node bifurcation. Final conclusions are presented in Section ~\ref{sec:conclusion}.
\section{Model description and basic results}\label{sec:fhnmodel}
In order to consider the effects of delay in the signal transmission between the neurons, we use the paradigmatic FHN model with quadratic term, \cite{fitzhugh1961impulses,nagumo1962active} and with time delay,  investigated by Wang et al. \cite{wang2009bifurcation}. 
 We here use a coupled nonidentical FHN neural system at which all the parameters are assumed to be non-negative:
 \begin{align}
   \nonumber\dot{v_{1}}(t)&= -v_{1}^{3}(t)+(a+1) v_{1}^2(t)-a v_{1}(t)-w_{1}(t)+ c \tanh (v_{2}(t-\tau)), \\
  \nonumber \dot{w_{1}}(t)&= \gamma v_{1}(t)- b_{1} w_{1}(t), \\
    \nonumber\dot{v_{2}}(t)&= -v_{2}^{3}(t)+(a+1) v_{2}^2(t)-a v_{2}(t)-w_{2}(t)+c \tanh (v_{1}(t-\tau)), \\
   \dot{w_{2}}(t)&= \gamma v_{2}(t)- b_{2} w_{2}(t), 
\label{1} 
 \end{align}
where $a, \gamma, b_{1}$ and $b_{2}$ are positive constants, $v_{1}$ represents the
membrane potential, $w_{1}$ is a recovery variable, $ c $ measures the coupling strength,  and $ \tau > 0 $ represents the time delay in signal transmission. We consider a quadratic term
as the self-connection function. We suppose that the function which describes the influence of a neuron on the other one, at time $t $, depends on the state of the neuron at some earlier time $t-\tau$. We consider $c$ and $ \tau $ as bifurcation parameters.
\subsection{Linear stability}
We can see that the origin is always a rest point of the system (\ref{1}), which we call the trivial equilibrium. The characteristic equation corresponding to linearization of the system (\ref{1}) at the trivial rest point is
 \begin{eqnarray} \label{2}
    P(c,\tau)&=& \lambda^{4}+ A \lambda^{3}+ B \lambda^{2}+ C \lambda + D \\
\nonumber- c^2 (\lambda&+&b_{1})(\lambda+b_{2}) e^{-2\lambda \tau}=0, 
\end{eqnarray}
where $ A=b_{1}+b_{2}+2a $, $ B=2a(b_{1}+2b_{2})+a^{2}+b_{1}b_{2}+2\gamma $, $ C= (a+b_{1})(ab_{2}+\gamma)+ (a+ b_{2})(ab_{1}+\gamma) $, and $ D= (ab_{1}+\gamma)(ab_{2}+\gamma)$.\\
Due to the presence of the delay, Eq. (\ref{2}) has infinitely many solutions; however, the stability of the equilibrium is determined by a finite number of critical roots with largest real parts.
\subsection{Occurrence of Hopf Bifurcation}
We want to obtain some conditions to ensure that the system (\ref{1}) undergoes a single Hopf bifurcation at the trivial rest point $(0,0,0,0)$, when $\tau$ passes through certain critical values.
 Substituting $ \lambda=i\omega$ into Eq. (\ref{2}) and separating the real and imaginary parts, we obtain
 \begin{align} \label{3}
\nonumber  \omega ^{4}&-B \omega^{2}+D-c^{2}(-\omega^{2} \cos(2\tau\omega)+b_{1}b_{2}\cos(2\tau\omega)\\
 \nonumber &+(b_{1}+b_{2})\omega sin(2\tau\omega))=0,\\
 \nonumber -A&\omega^{3}+C\omega-c^{2} (\omega^{2} \sin(2\tau\omega)-b_{1}b_{2}\sin(2\tau\omega)\\
 & + (b_{1}+b_{2})\omega cos(2\tau\omega))=0.
\end{align} 
Eliminating $\tau$ from Eq. (\ref{3}) gives  
 \begin{eqnarray}  \label{6}
\resizebox{.45\hsize}{!}{$  \omega^{8}+P \omega^{6}+Q \omega^{4}+R \omega^{2}+S=0$},
\end{eqnarray}
where $  P=-2B+A^{2} $, $ Q=B^{2}+2D-2AC-c^{4} $, $ R=-2BD+C^{2}-c^{4}(b_{1}^{2}+b_{2}^{2}) $, and $S= D^{2}-c^{4}(b_{1}b_{2})^{2}$. For $ z=\omega^{2} $, we get
 \begin{eqnarray}\label{4} 
   z^{4}+P z^{3}+Q z^{2}+R z+S=0. 
\end{eqnarray}
Since the form of Eq. (\ref{4}) is identical to that of Eq. (2.4) in the paper of Li and Wei \cite{li2005zeros}, we may apply Lemmas 2.2 and 2.3 analogously.
Denote $h(z)=z^{4}+P z^{3}+Q z^{2}+R z+S$, then we have $h'(z)= 4 z^{3}+3P z^{2}+2Q z+R$. Set $4 z^{3}+3P z^{2}+2Q z+R=0$. Let $y=z+\frac{3P}{4}$, then the Eq. (\ref{4}) becomes $y^{3}+P_{1}y+Q_{1}=0$, where $P_{1}= \frac{Q}{2}-\frac{3}{16}P^{2}$ and $ Q_{1}=\frac{P^{3}}{32}-\frac{PQ}{8}+\frac{R}{4}$. Define $\Delta=(\frac{Q_{1}}{2})^{2}+(\frac{P_{1}}{3})^{3}$, $\epsilon=\frac{-1+i\sqrt{3}}{2}$,
 \begin{eqnarray}\label{5} 
 \nonumber y_{1}&=&{\textstyle \sqrt[3]{-\frac{Q_{1}}{2}+\sqrt{\Delta}}+ \sqrt[3]{-\frac{Q_{1}}{2}-\sqrt{\Delta}}}, \\
 \nonumber y_{2}&=&{\textstyle\sqrt[3]{-\frac{Q_{1}}{2}+\sqrt{\Delta}}\epsilon+ \sqrt[3]{-\frac{Q_{1}}{2}-\sqrt{\Delta}}}\epsilon^{2}, \\
 \nonumber y_{3}&=&{\textstyle\sqrt[3]{-\frac{Q_{1}}{2}+\sqrt{\Delta}}\epsilon^{2}+ \sqrt[3]{-\frac{Q_{1}}{2}-\sqrt{\Delta}}}\epsilon. 
 \end{eqnarray}
Let $z_{i}=y_{i}-\frac{P}{4}$, ($i$=1,2,3).
Suppose that Eq. (\ref{4}) has positive roots. Without loss of generality, we assume that it has four positive roots, denoted by $z^{\ast}_{k}  (k=1,2,3,4)$. Then Eq. (\ref{6}) has four positive roots, say $\omega_{i}=\sqrt{z^{\ast}_{i}}$, $i=1,2,3,4$. By Eq. (\ref{3}) we have
\small
 \begin{eqnarray}\label{7}
\nonumber  \sin(2\tau \omega)=\frac{(\omega_{k}^{4}-B\omega_{k}^{2}+D)(b_{1}+b_{2})\omega_{k}}{c^{2}[(b_{1}+b_{2})^{2}\omega_{k}^{2}+(\omega_{k}^{2}-b_{1}b_{2})^{2}]}\\
\nonumber + \frac{(A\omega_{k}^{3}-C\omega_{k})(-\omega_{k}^{2}+b_{1}b_{2})}{c^{2}[(b_{1}+b_{2})^{2}\omega_{k}^{2}+(\omega_{k}^{2}-b_{1}b_{2})^{2}]},\\
\nonumber \cos(2\tau \omega)=\frac{(\omega_{k}^{4}-B\omega_{k}^{2}+D)(b_{1}b_{2}-\omega^{2}_{k}}{c^{2}[(b_{1}+b_{2})^{2}\omega_{k}^{2}+(\omega_{k}^{2}-b_{1}b_{2})^{2}]}\\
+\frac{(-A\omega_{k}^{3}+C\omega_{k})((b_{1}+b_{2})\omega_{k})}{c^{2}[(b_{1}+b_{2})^{2}\omega_{k}^{2}+(\omega_{k}^{2}-b_{1}b_{2})^{2}]}.
 \end{eqnarray}
\normalsize
Thus, denoting
\small
 \begin{eqnarray} \label{astar}
\nonumber a^{\ast}=\frac{(\omega_{k}^{4}-B\omega_{k}^{2}+D)(b_{1}+b_{2})\omega_{k}}{c^{2}[(b_{1}+b_{2})^{2}\omega_{k}^{2}+(\omega_{k}^{2}-b_{1}b_{2})^{2}]}\\
\nonumber + \frac{(A\omega_{k}^{3}-C\omega_{k})(-\omega_{k}^{2}+b_{1}b_{2})}{c^{2}[(b_{1}+b_{2})^{2}\omega_{k}^{2}+(\omega_{k}^{2}-b_{1}b_{2})^{2}]},\\
\nonumber b^{\ast}=\frac{(\omega_{k}^{4}-B\omega_{k}^{2}+D)(b_{1}b_{2}-\omega^{2}_{k}}{c^{2}[(b_{1}+b_{2})^{2}\omega_{k}^{2}+(\omega_{k}^{2}-b_{1}b_{2})^{2}]}\\
 +\frac{(-A\omega_{k}^{3}+C\omega_{k})((b_{1}+b_{2})\omega_{k})}{c^{2}[(b_{1}+b_{2})^{2}\omega_{k}^{2}+(\omega_{k}^{2}-b_{1}b_{2})^{2}]},
 \end{eqnarray}
 \begin{eqnarray}\label{tau}
\nonumber \tau^{(j)}_{k}&=\left\{
\begin{array}{ll}
   \frac{1}{2\omega_{k}}(\arccos b^{\ast}+2j\pi),& a^{\ast}\geqslant 0,\\
    \frac{1}{2\omega_{k}}(2\pi -\arccos b^{\ast}+2j\pi),& a^{\ast}< 0,
\end{array} \right.\\
 &(k=1,2,3,4, j=0,1,2,...)
  \end{eqnarray}
\normalsize
then $\pm i\omega_{k} $ is a pair of purely imaginary roots of (\ref{2}) with $\tau=\tau^{(j)}_{k}$.\\
To facilitate the reading, Lemmas and Theorems about roots of characteristic equation,  Eq. (\ref{2}), and possible steady state bifurcations of the trivial rest point are presented in Appendix. \\
In the rest of the paper we will study the dynamical structure of the system. We will explain coupling and delay driven dynamics of the system (\ref{1}) by considering possible bifurcations. Bifurcation diagrams are obtained numerically or analytically from the mathematical model, and the scrutiny of bifurcation diagrams clarifies the parameter regions of different behaviors. Numerical simulations using the bifurcation analysis software DDE-Biftool \cite{engelborghs2002numerical} are carried out to illustrate the main results and to support the analysis of our system.
\section{Trivial equilibrium}\label{sec:trivial equilibria}
We want to consider possible rest points and limit cycles of the system (\ref{1}), and their bifurcations for the variable parameters $c$ and $\tau$'s. In this section we will focus on bifurcations for values of coupling strength not much bigger than one.  \\
 As stated in the previous section the origin is always a rest point of the system, without any constraint on the parameters.
We should emphasize that each neuron without coupling is in silent mode. It isn't hard to see that for $\frac{b_{i}}{\gamma}<\frac{4}{(a-1)^{2}}$, $i=1 ,2$ the origin is the unique equilibrium of each single neuron. We use a variable change, to transform the system to a special type of Lienard system, as $x=-v_{i}$ and $y=w_{i}+b_{i}x$. This yields the Lienard system $\dot{x}=y-F(x)$ and $\dot{y}=-g(x)$, where $F(x)=x^{3}+(a+1)x^{2}+(a+b_{i})x$ and $g(x)=b_{i}(x^{3}+(a+1)x^{2}+(a+\frac{\gamma}{b_{i}}))$, $i= 1 , 2$. We can see that the conditions of Theorem 1 of \cite{ringkvist2005existence} are satisfied, for $\frac{b_{i}}{\gamma}<\frac{4}{(a-1)^{2}}, i=1 , 2$, and the single neurons without coupling have no limit cycles. Also for $b_{i}>\frac{a^{2}-a+1}{3}$ the $div=-3v_{i}^{2}+2(a+1)v_{i}-a-b, i=1 , 2$, is negative and by Bendixson-Dulac theorem, it can be seen that the single neuron of the FHN model does not admit periodic solutions. If we choose parameter $a=0.3$, we can see that for $\gamma>\frac{1}{30}$ at least one of the above stated inequalities are satisfied and single neurons have no limit cycles for arbitrary value of parameter $b$.
  In the paper we choose the value of the parameters as $a=0.3$, $\gamma=0.3$, $b_{1}=0.15$, and $b_{2}=0.18$. It is easy to check that for these values of parameters the above stated arguments are satisfied and each single neuron without coupling is at rest.\\
   For our system we can see that when $c=0$ the origin, $M_{0}=(0,0,0,0)$, as a trivial equilibrium is always stable. First, we want to study the bifurcations of the trivial equilibrium according to coupling strength and time delay.
\subsection{Hopf Bifurcations of $M_{0}$ }
\begin{figure}
\centering
\resizebox{0.7\textwidth}{!}{
\includegraphics{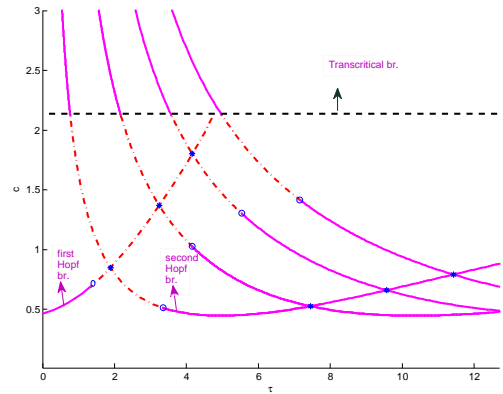}
}
\vspace{-1.5cm}
\caption{\footnotesize Hopf branches of trivial equilibrium. Dash-dot red lines are sub-critical Hopf bifurcation branches. Solid magenta lines are super-critical Hopf branches. The dashed black line is the branch of transcrirical bifurcation. Circles and stars show Bautin and double-Hopf bifurcations respectively.}
\label{total}
\end{figure}

\begin{figure}
\centering
\resizebox{0.7\textwidth}{!}{
\includegraphics{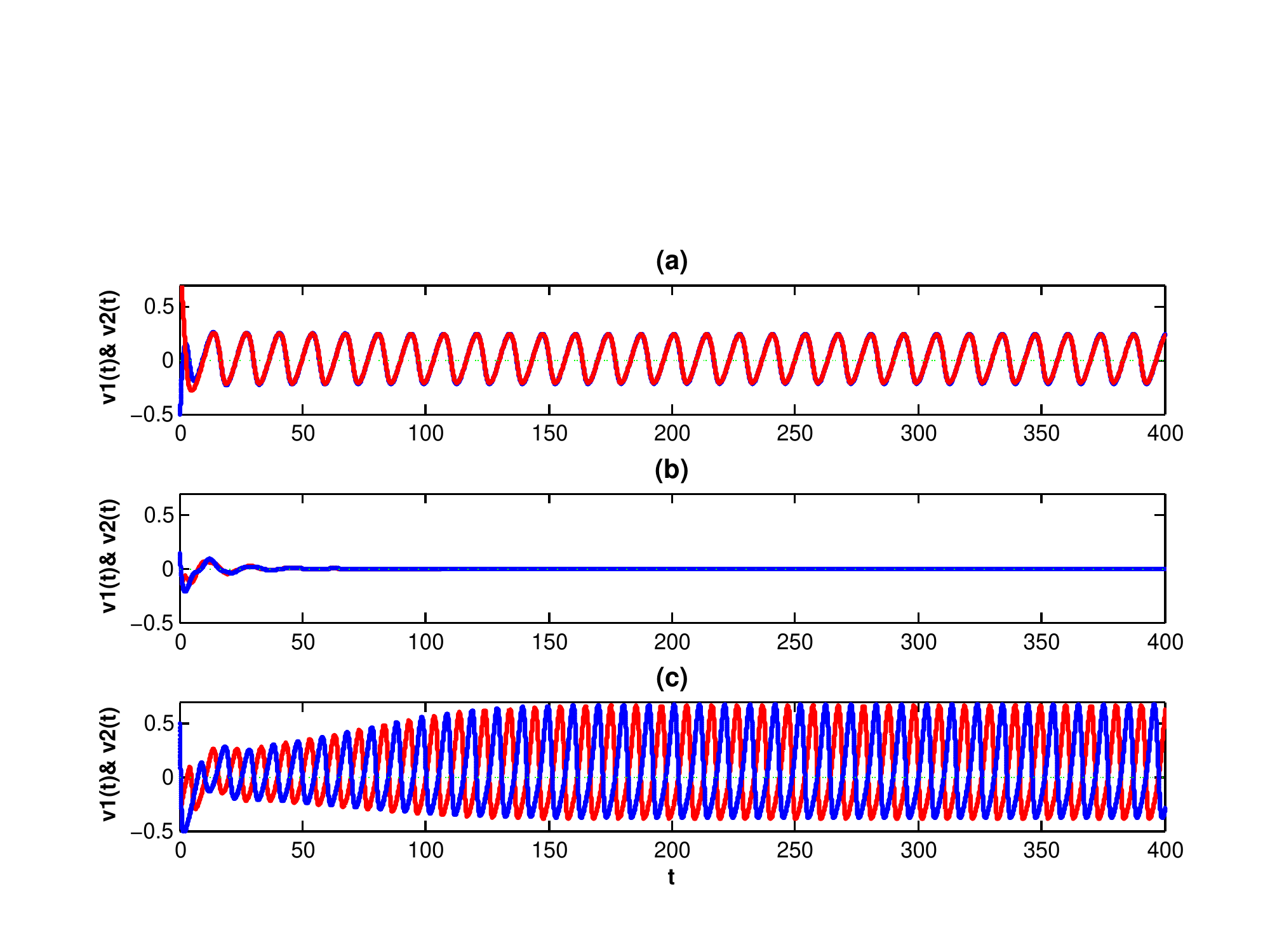}
}
\vspace{-0.4cm}
\caption{\footnotesize Different modes of spiking due to different $\tau$'s, here $c=0.5$. a) $\tau$=0.2, b) $\tau$=2, c) $\tau$=4.}
\label{modes}
\end{figure}

We start the study of possible bifurcations of the trivial rest point with finding branches of Hopf bifurcation.
It is not hard to see that for parameters $c\leq0.447$, the conditions of Lemma \ref{stability} hold, and the origin is the only stable rest point for $\tau\geq 0$. For the system (\ref{1}) neurons are excitable because the trivial resting state is near a Hopf bifurcation, i.e., near a transition from quiescence to spiking. Hence, by varying $c$ and $\tau$, a branch of Hopf bifurcation emanates from $c\simeq0.4646$ and $\tau=0$, in $(c,\tau)$ plane. By changing the parameters $c$ and $\tau$ the other Hopf branches emanates, see Fig. \ref{total}.
In order to understand the bifurcation diagram of Fig. \ref{total}, we can fix the parameter $c$ and study the dynamics of the system according to $\tau$. As it is shown in Fig. \ref{modes}(a), for $c=0.5$ and $\tau=0.2$ the neurons oscillate synchronously,
in which the phase shift between oscillation of two neurons is zero. By increasing the parameter $\tau$ the amplitude
of the periodic solution decreases until the parameter $\tau$ reaches the first branch of Hopf bifurcation for $\tau=\tau_{0}$, where $\tau_{0}= \tau^{(0)}_{k}$, according to Eq. (\ref{tau}). Due to the super-critical Hopf bifurcation the stable limit cycle disappears, and the trivial rest point becomes stable. Therefore for $\tau\in(\tau_{0},\tau_{1})$, where according to Eq. (\ref{tau}), $\tau_{1}= \tau^{(1)}_{k}$, the only stable state of the system is the trivial rest point, Fig. \ref{modes}(b). By further increasing the parameter $\tau$, the neurons start to spike in an anti-phase manner, in which the phase shift between oscillation of two neurons is equal to $\pi$, Fig. \ref{modes}(c). This is due to the second branch of super-critical Hopf bifurcation. By increasing the range of parameter $\tau$ other branches of Hopf bifurcation appear, and the above stated scenario repeats.
When we want to study the impact of Hopf bifurcations for larger values of $c$ or $\tau$, we can see that in Hopf bifurcation branches there are Bautin bifurcation points, which separates branches of sub- and super-critical Hopf bifurcation curves in the parameter plane. Due to the Bautin bifurcation a branch of fold of limit cycles bifurcation emerges from the Bautin point. These fold of limit cycle branches are close to sub-critical Hopf branches. Also, there are some points in the bifurcation diagram of the trivial rest
point, Fig. \ref{total}, which can be obtained from the intersection of the two branches of Hopf bifurcation. It is possible that the characteristic equation at a rest point has eigenvalues with strict nonzero real parts except two pairs of purely imaginary eigenvalues; in $(\tau,c)$ plane it happens when two branches of Hopf bifurcation cross each other at a point, this situation is called a double-Hopf bifurcation. In the next section we want to describe the dynamics in two different areas. First, in the regions which lies above two first Bautin points and below the first double-Hopf point. Second, in the regions around the first double-Hopf point.
\subsubsection{Bautin Bifurcations}
As stated in the previous section there are fold of limit cycles branches which emanates from Bautin Bifurcation points. To understand these branches and to study the impact of parameter $\tau$, we can fix the parameter $c$ above the first two Bautin bifurcation points, and study the dynamic changes according to $\tau$.
\begin{figure}
\centering
\resizebox{0.7\textwidth}{!}{
\includegraphics{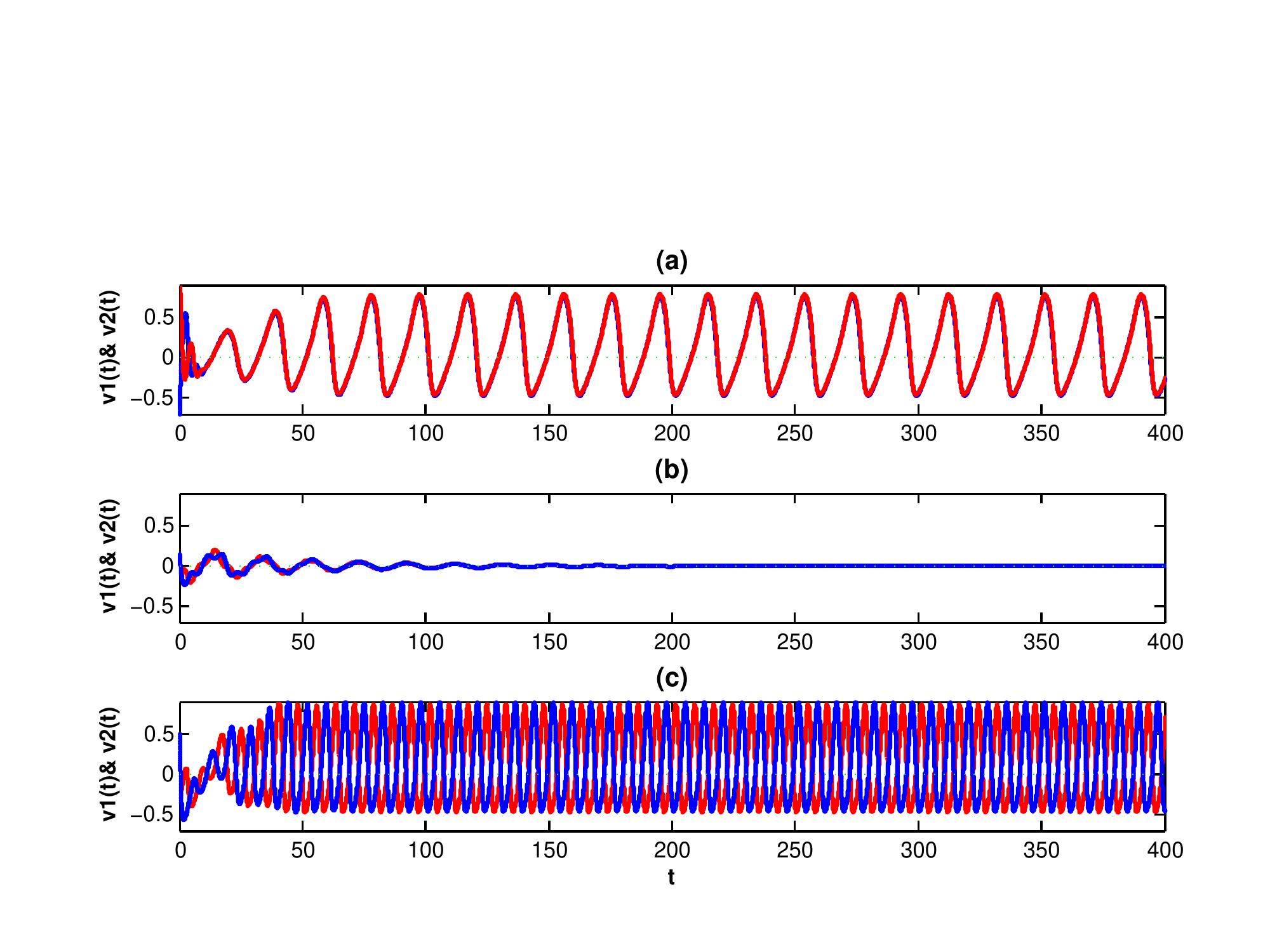}
}
\vspace{-0.4cm}
\caption{\footnotesize Different modes of spiking due to different $\tau$'s, here $c=0.8$. a) $\tau$=1.2, b) $\tau$=1.9, c) $\tau$=1.6.}
\label{abovebautin}
\end{figure}
As it is shown in Fig. \ref{abovebautin}(a), for $c=0.8$ and $\tau=0.2$ (region I in  Fig. \ref{doublehopf} ), the neurons oscillate synchronously. By a little increase of the parameter $\tau$ the parameter reaches the first branch of Hopf bifurcation which is sub-critical. Due to the sub-critical Hopf bifurcation an unstable limit cycle with index 1 appears and the trivial equilibrium $M_{0}$ which was unstable with index 2 becomes stable.
By a little increase of the parameter $\tau$ the amplitude of the periodic solution decreases until the parameter $\tau$ reaches the first branch of fold of limit cycles bifurcation. Due to the fold of limit cycles bifurcation the stable and unstable with index 1 limit cycles collide and disappear. The only stable state of the system (region II of Fig. \ref{doublehopf}) is the trivial rest point, Fig. \ref{abovebautin}(b). By further increasing the parameter $\tau$, the second branch of fold of limit cycles bifurcation appears and two limit cycles emerge. One of them is stable and the other one is unstable with index 1. Therefore the neurons start to spike in an anti-phase manner, Fig. \ref{abovebautin}(c). By a little increase of the parameter $\tau$ the second branch of Hopf bifurcation is met (region III in Fig. \ref{doublehopf}). Due to the sub-critical Hopf bifurcation the unstable limit cycle with index 1 disappears, and the trivial rest point becomes unstable. We should notice that the second fold and Hopf branches are purely delay-driven and have no trace in system with instantaneous coupling. Moreover, there is an interesting observation about anti-phase and synchronized solutions. Actually, synchronized and anti-phase activities of the coupled neurons can be achieved in some parameter ranges related to their bifurcation transition. For an explanation about the mechanism of these kind of transition between synchronized and anti-phase solutions see \cite{FarajzadehTehrani201541}.  
 By increasing the range of parameter $\tau$ other branches of Hopf and corresponding fold of limit cycles bifurcations appear.  
 Also we can see that for larger values of parameter $\tau$ there are branches of torus bifurcation. 
  If we consider other fold and torus branches for larger values of parameter $\tau$, we can see that there are regions in which multistability occurs in the system.\\
\begin{figure}
\begin{tabular}{cc}
\resizebox{0.55\textwidth}{!}{
\includegraphics{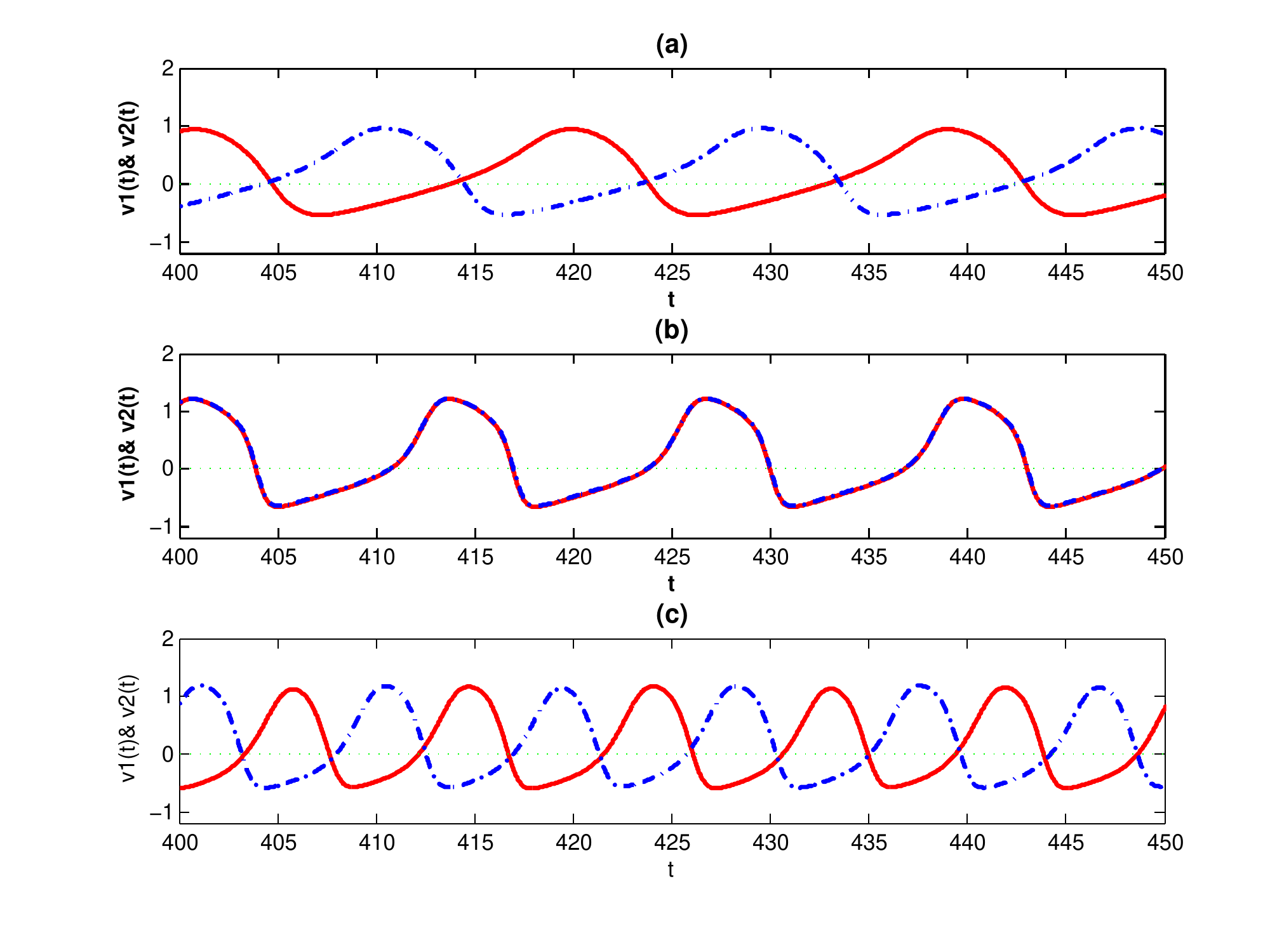}
}
\resizebox{0.55\textwidth}{!}{
\includegraphics{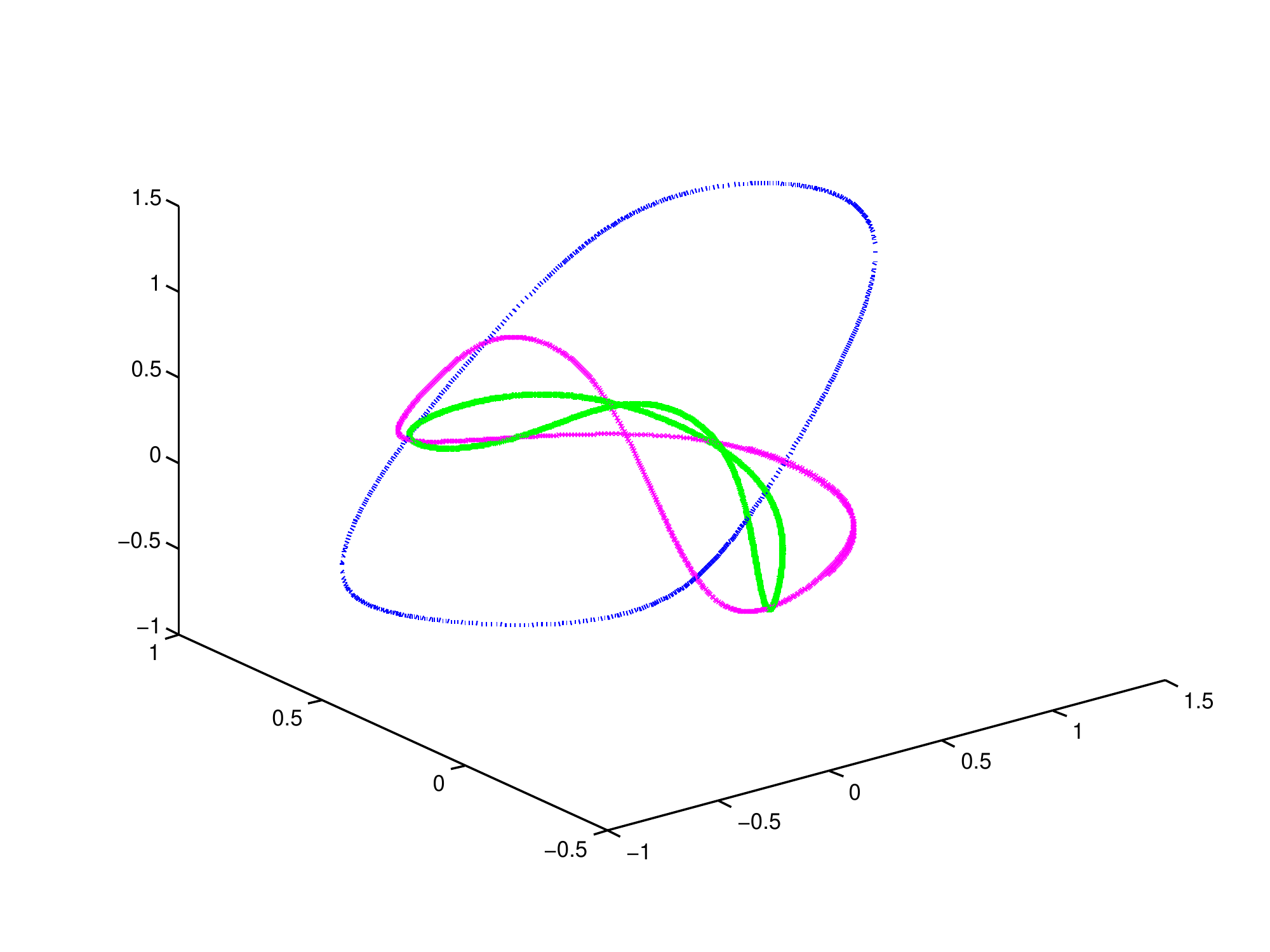}
}

\end{tabular}
\caption{\footnotesize Left: Different modes of spiking due to different initial conditions. Right: The projection of different periodic solutions, depicted in Left figure, on $(v_{1},w_{1},v_{2})$ plane. The fixed parameters are $c=0.8$ and $\tau=10.5$.}
\label{underdoublehopf}
\end{figure}  
   In the previous paragraph we analyzed the changes of dynamic for $c=0.8$, and a range of parameter $\tau$. If we increase the parameter $\tau$, another branch of fold of limit cycles appears and generates two limit cycles with index 2 and 3. The limit cycle with index 3 disappears through Hopf bifurcation. By increasing the parameter $\tau$ the limit cycle with index 2 becomes stable through torus bifurcation. Therefore, For a range of parameter $\tau$, for example $\tau=6.8$, there are two stable limit cycles. By increasing the parameter $\tau$ another fold of limit cycles bifurcation and two torus bifurcations occur. As a result of these bifurcations three stable limit cycles exist simultaneously in a region of parameters.
 As an example when $\tau=10.5$, there are three different periodic solutions with different periods, which are depicted in Fig. \ref{underdoublehopf}.
We should emphasize that finding all of these periodic solutions by numerical simulation is very difficult, and requires analytical investigations.\\
\textbf{Remark}. We should notice that, even for small delays, the parameter $\tau$ can either suppress periodic spiking or induce new periodic spiking, depending on the value of the time delay, see Fig. \ref{abovebautin}. Thus, treatment of neural systems with changing the time delay can be delicate and challenging.
\subsubsection{Double-Hopf Bifurcation}
\begin{figure}
\centering
\resizebox{1.0\textwidth}{!}{
\includegraphics{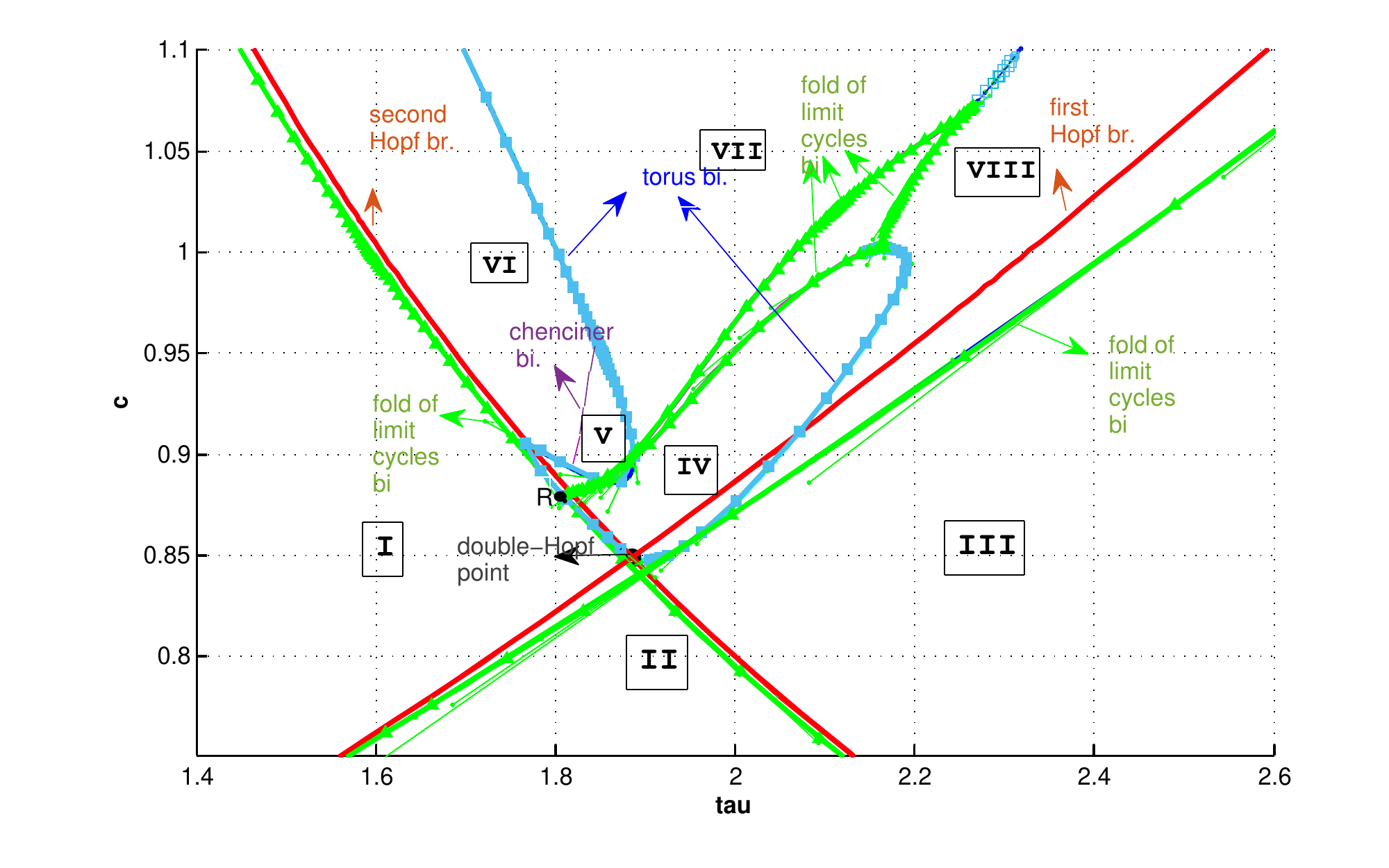}
}
\vspace{-0.7cm}
\caption{\footnotesize Bifurcation diagram of trivial equilibrium around the first double-Hopf point.}
\label{doublehopf}
\end{figure}

\begin{figure}
\centering
\resizebox{0.6\textwidth}{!}{
\includegraphics{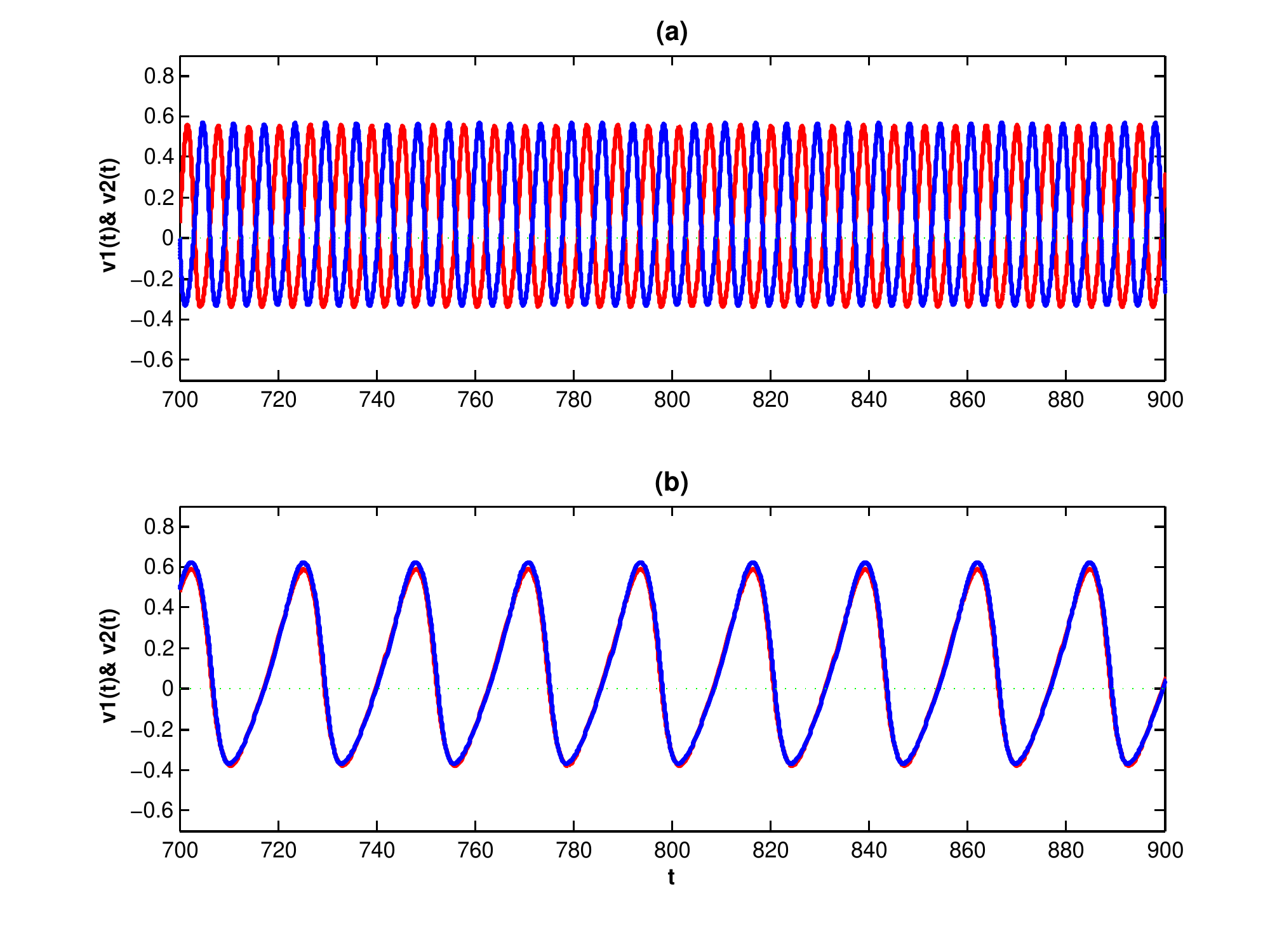}
}
\vspace{-0.4cm}
\caption{\footnotesize Bi-stability in region IV. Two different modes of spiking due to different initial conditions, $c=0.865$, $\tau=1.9$.}
\label{regionIV}
\end{figure}

\begin{figure}
\begin{tabular}{cc}
\resizebox{0.5\textwidth}{!}{
\includegraphics{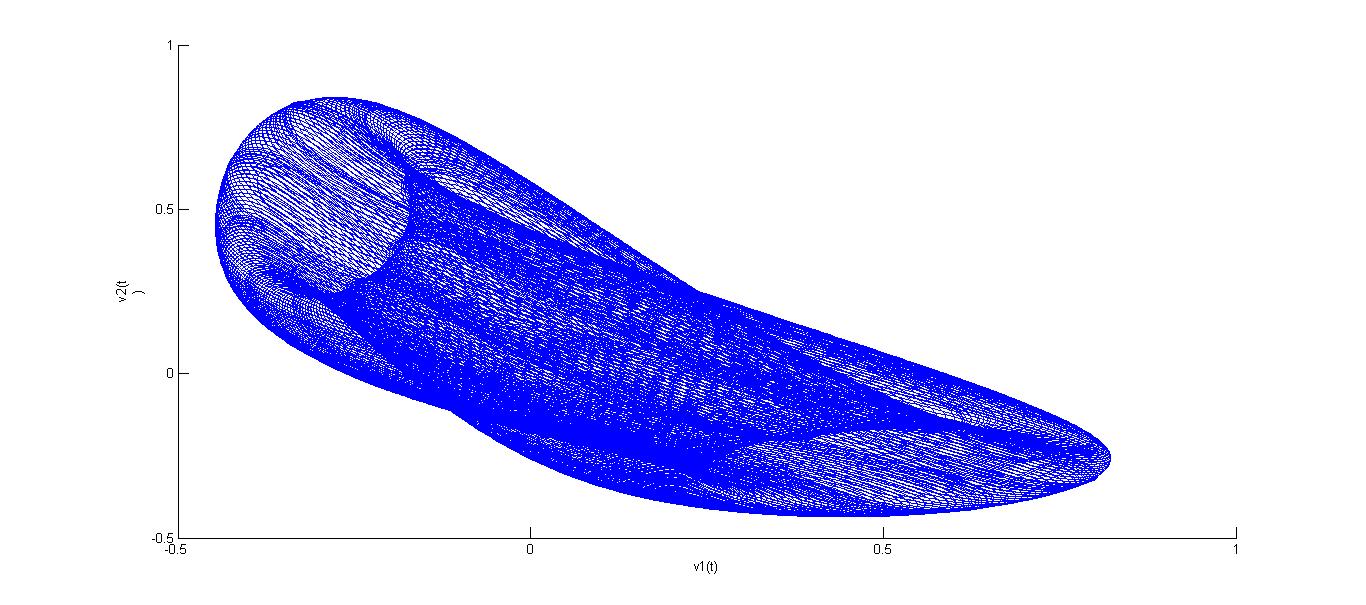}
}
\resizebox{0.5\textwidth}{!}{
\includegraphics{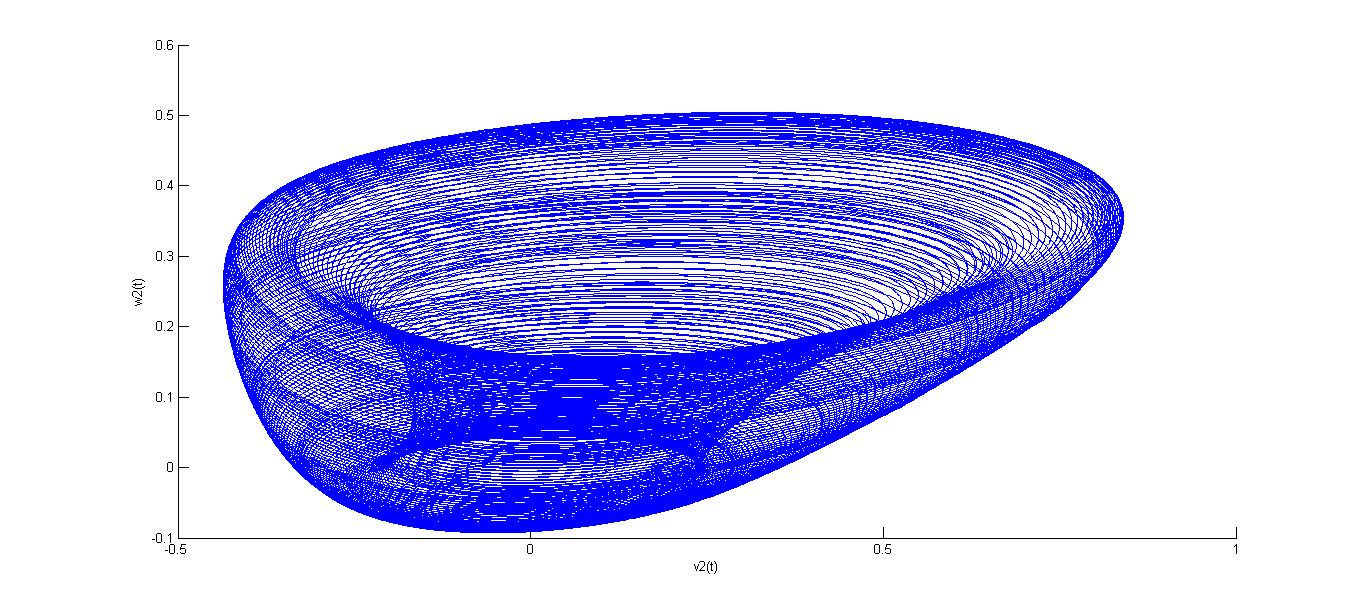}
}\\
\resizebox{0.5\textwidth}{!}{
\includegraphics{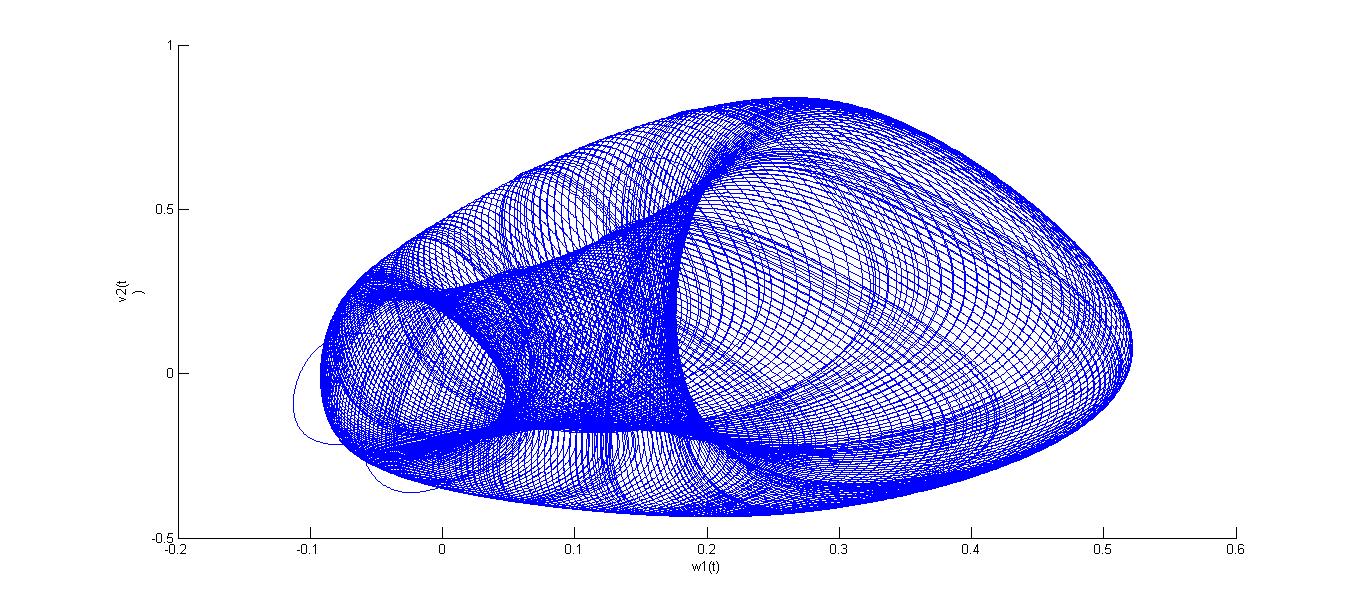}
}
\end{tabular}
\caption{\footnotesize Projection of stable torus on two dimensional planes.}
\label{v1w1v2}
\end{figure}

\begin{figure}
\centering
\resizebox{0.65\textwidth}{!}{
\includegraphics{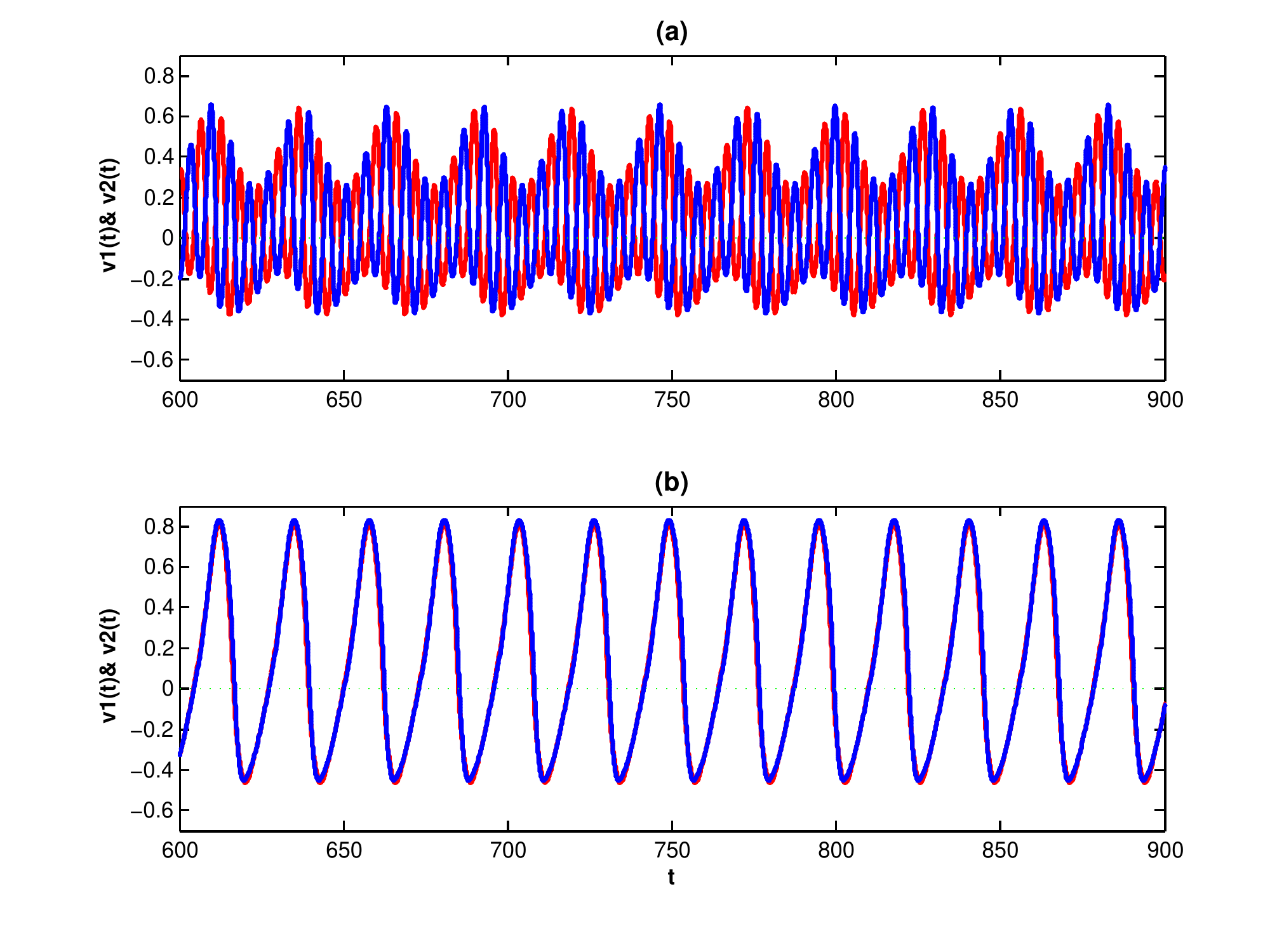}
}
\vspace{-0.7cm}
\caption{\footnotesize Bi-stability in region V. a) Stable torus, b) stable limit cycle.}
\label{regionV}
\end{figure}

We want to focus on the first two Hopf bifurcation branches which intersect and make double-Hopf point. In the previous section we explained the bifurcations and dynamics under this point. Now we will discuss the dynamics around this double-Hopf point, see Fig. \ref{doublehopf}. We can see that the dynamics near other double-Hopf points is similar.
We want to start the study of dynamics around double-Hopf point in the regions below this point (for $c<0.849$). For $c=0.8$, the dynamic changes between regions I, II, and III of Fig. \ref{doublehopf}, is studied in the previous section.
We increase the parameter $c$ such that double-Hopf bifurcation occurs. Above the double-Hopf bifurcation we fix the parameter $c=0.86$ and change the parameter $\tau$ between regions I, IV, and III, see Fig. \ref{doublehopf}. During passing from region I to IV, two limit cycles ( stable and unstable with index 1) appear through fold bifurcation of limit cycles. By a little change of parameter $\tau$, the index-1 unstable limit cycle changes to index-3 unstable limit cycle through a torus bifurcation. The index-3 unstable limit cycle disappears via a sub-critical Hopf bifurcation. Therefore in region IV the system is bi-stable, one stable limit cycle which corresponds to synchronized solutions and another stable limit cycle which appeared through fold cycle bifurcation and is correspondent to anti-phase solutions, Fig. \ref{regionIV}.
 By increasing the parameter $\tau$ and during passing from region IV to III, an unstable limit cycle with index 3 appears through sub-critical Hopf bifurcation. By a little
increase of parameter $\tau $ the index-3 unstable limit cycle changes to index-1 unstable limit cycle through torus bifurcation. If we increase the parameter $\tau$ a little bit more, the index-1 limit cycle and the stable limit cycle which is correspondent to synchronized solutions collide and disappear through fold bifurcation. As stated in the previous paragraph, in region III the only stable state of the system is the limit cycle which is correspondent to anti-phase solutions. \\
In Fig. \ref{doublehopf} there is a point marked by $R$, in which two branches of fold cycle bifurcation and a branch of torus bifurcation meet each other. For such values of parameters the limit cycle on which torus bifurcation occurs has simple  floquet multipliers $\pm i$. This bifurcation is known as $1:4$ strong resonance, \cite{kuznetsov2013elements}.\\
We should notice that there is a point in both  branches of fold bifurcation of limit cycles in which torus and fold cycle branches become tangent to each other. This is a fold-torus bifurcation \cite{kuznetsov2013elements,guckenheimer1983nonlinear}, or in other words, fold-Neimrak-sacker bifurcation in Poincare section \cite{broer2008hopf,broer2005hopf}. This bifurcation occurs when a complex-conjugate pair
of floquet multipliers and a floquet multiplier equal one cross the unit circle. In one side of this point the fold bifurcation of limit cycles produces two limit cycles with index 0 and 1, and in the other side fold bifurcation creates two limit cycles with index 2 and 3. Therefore during entering from region I to VI, two limit cycles with index 2 and 3 appear. The index-3 limit cycle disappears through Hopf bifurcation and index-2 limit cycle remains unchanged. Therefore in region VI the only stable state of the system is the limit cycle which is correspondent to synchronized solutions. \\
As parameters $c$ and $\tau$ change from region VI to V, two tori appear through Chenciner (generalized Neimark-Sacker) bifurcation \cite{kuznetsov2013elements}, one stable and one unstable, the stable one is depicted in Fig. \ref{v1w1v2}.
 Therefore the system in region V is bi-stable; a torus and a limit cycle, Fig. \ref{regionV}. 
Also we should notice that in the points in which Chenciner branch meets torus branches the sub-critical and super-critical tori branches separate from each other, \cite{kuznetsov2013elements}. By changing parameters from region V to VII the super-critical torus bifurcation occurs and the stable torus disappears and makes the limit cycle stable. In result, in region VII the system is bi-stable with two limit cycles.
\subsection{Multistability}\label{sec:multistability}
\begin{figure}
\centering
\resizebox{0.7\textwidth}{!}{
\includegraphics{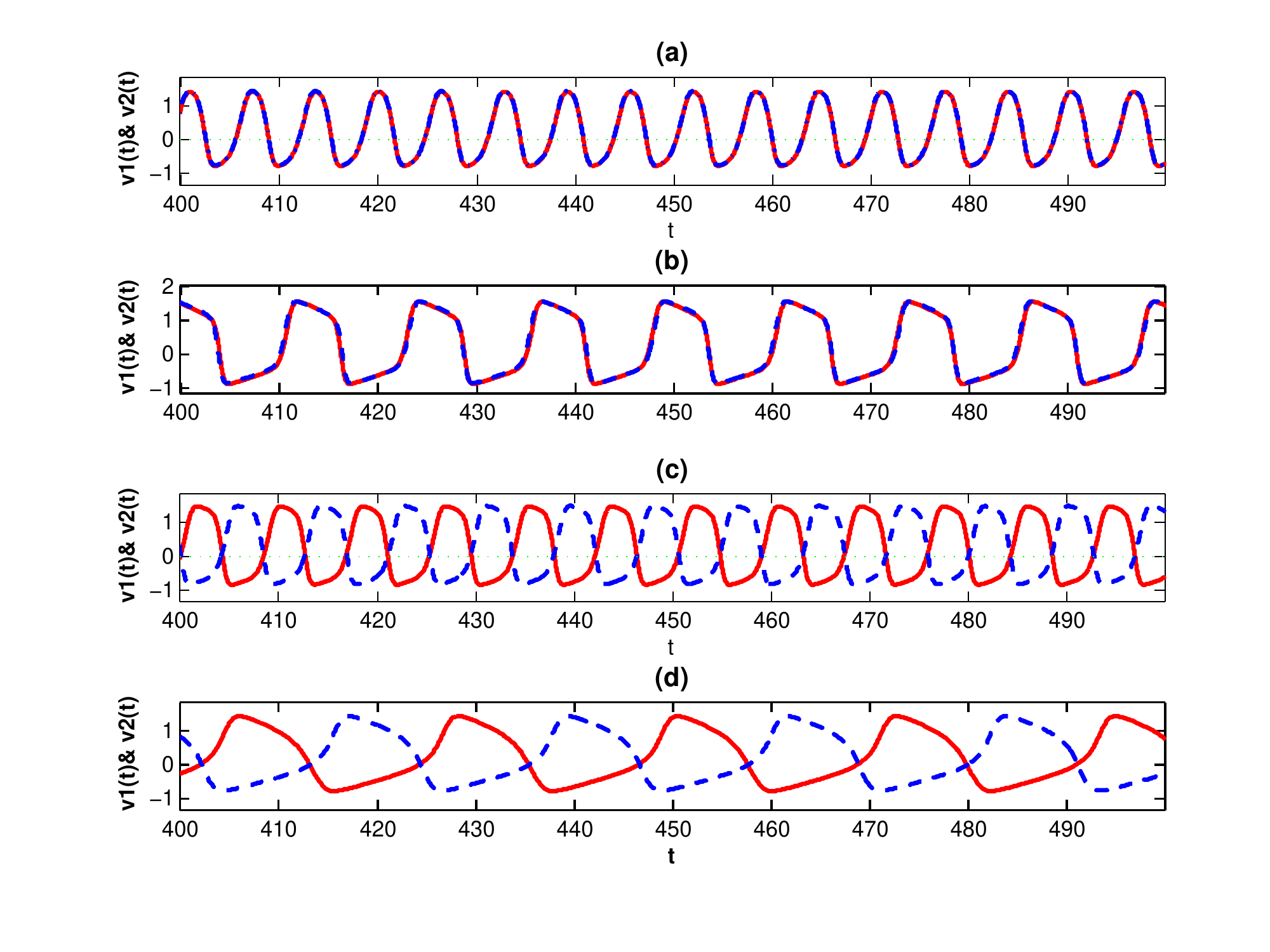}
}
\vspace{-0.7cm}
\caption{\footnotesize Different modes of spiking due to different initial conditions, here $c=1.5$, and $\tau\simeq 12$. }
\label{4stabletimeseries}
\end{figure}

\begin{figure}
\centering
\resizebox{0.7\textwidth}{!}{
\includegraphics{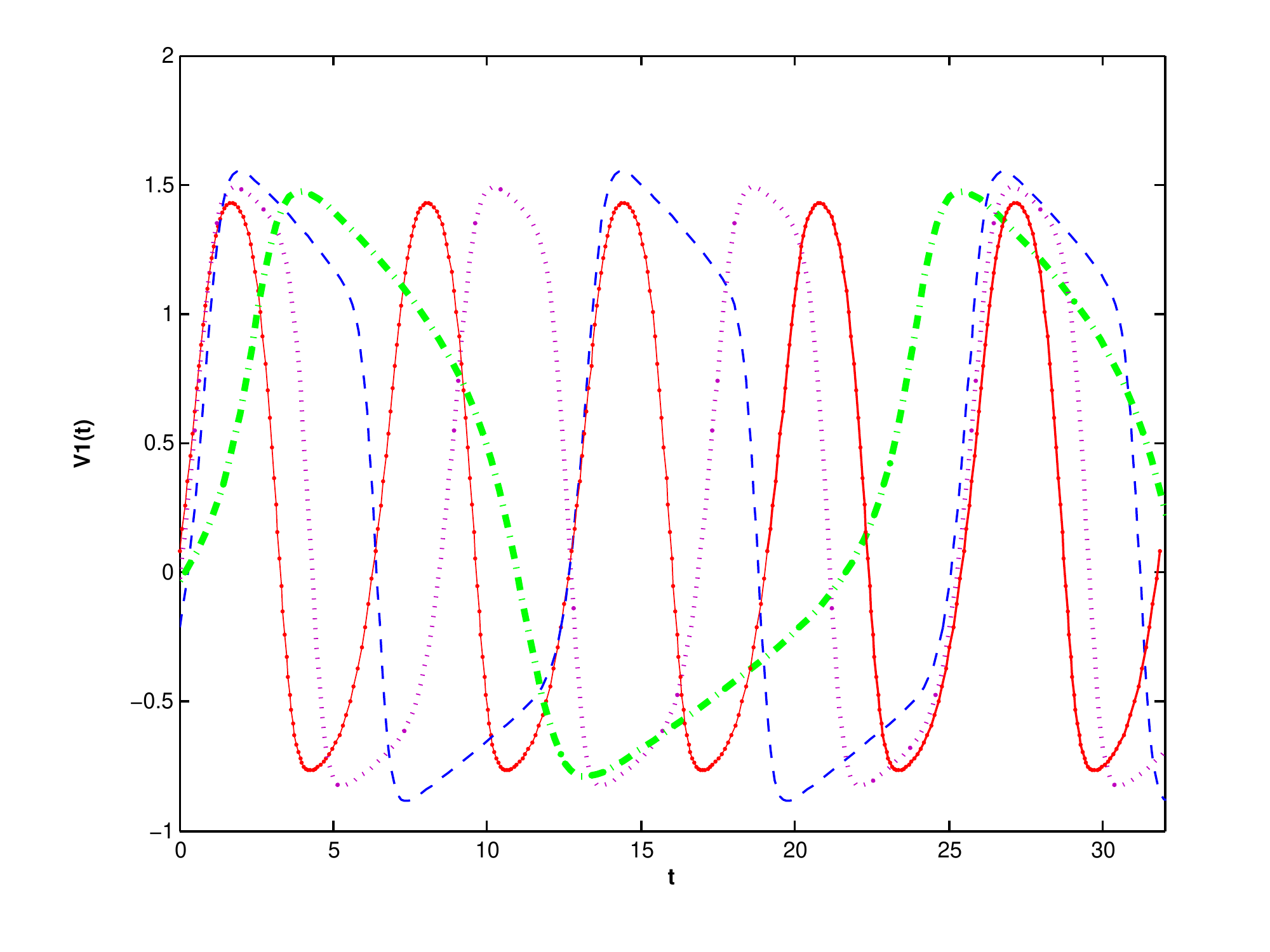}
}
\vspace{-0.7cm}
\caption{\footnotesize The dash-dot green line is correspondent to anti-phase activities, Fig. \ref{4stabletimeseries}(d). The dashed blue line is correspondent to synchronized activities, Fig. \ref{4stabletimeseries}(c). The doted magenta line is correspondent to anti-phase activities, Fig. \ref{4stabletimeseries}(b). The solid red line with circle marker is correspondent to synchronized activities, Fig. \ref{4stabletimeseries}(a). }
\label{4stablepsol}
\end{figure}

In the previous section we analyzed the activities around two first branches of fold of limit cycles bifurcations and their corresponding torus bifurcations. If we consider other fold and torus branches for larger values of parameter $\tau$, we can see that there are regions in which multistability occurs in the system. For example, if we fix the parameter $c=1.5$, by increasing the parameter $\tau$ several fold of limit cycles and torus bifurcations occur. As a result of these bifurcations several stable limit cycles appear. As an example when $\tau \simeq 12$, there are four different periodic solutions, Fig. \ref{4stabletimeseries}. 
The stated limit cycles have different periods, Fig. \ref{4stablepsol}. Two of them are correspondent to synchronized activities, and the other two are correspondent to anti-phase activities.
We should emphasize that finding all these periodic solutions by numerical simulation is very difficult. Actually we found them by following the bifurcations of periodic solutions.
\section{Strong coupling}\label{sec:strong}
\begin{figure}
\centering
\resizebox{0.9\textwidth}{!}{
\includegraphics{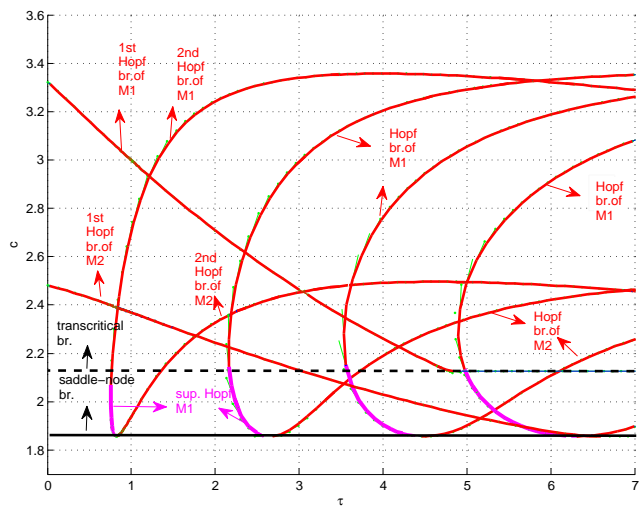}
}
\vspace{-0.9cm}
\caption{\footnotesize Hopf bifurcation branches of $M_{1}$ and $M_{2}$ equilibria. Red lines show sub-critical Hopf bifurcations. Magenta lines show super-critical Hopf bifurcations of $M_{1}$ equilibrium. }
\label{hopfm1m2}
\end{figure}

Now we want to increase the parameter $c$ and study the changes of dynamics for strong coupling of two neurons. In the next sections we will see that the dynamics specially in stable states are different for strong coupling of two neurons.
We should emphasize that by considering strong coupling we mean that each neuron is representative of a network of neurons.\\
We know that $M_{0}=(0,0,0,0)$ is always a rest point of the system (\ref{1}). In addition to $M_{0}$ we found that for fixed parameters $a=0.3$, $b_{1}=0.15$, $b_{2}=0.18$, $\gamma=0.3$, and for strong coupling, there are two non-trivial rest points $M_{1}$ and $M_{2}$.  Actually for the parameter $c\simeq1.858$, the saddle-node bifurcation occurs and two non-trivial equilibria $M_{1}$ and $M_{2}$ appear, which are both in positive orthant, Fig. \ref{hopfm1m2}. In order to study the impact of $M_{1}$ and $M_{2}$ on the dynamics of the system, first we consider Hopf bifurcations of these equilibria, see Fig. \ref{hopfm1m2}. As it is depicted in Fig. \ref{hopfm1m2}, there is a saddle-node branch such that the Hopf branches of $M_{1}$ and $M_{2}$ stop when reach it. It implies that for parameters $c<1.858$ the origin is the only equilibrium of the system.
 The $M_{2}$ equilibrium stays always on positive orthant, and for parameters $c>2.481$ and arbitrary $\tau$ is always stable. The $M_{1}$ equilibrium at first is in the positive orthant, but through transcritical bifurcation ($c\simeq2.12$) interchanges its stability with trivial rest point $M_{0}$, and enters the negative orthant and stays there for larger values of parameter $c$. Also as it is depicted in Fig. \ref{hopfm1m2}, transcritical branch separates sub- and super-critical Hopf branches of $M_{1}$ from each other. We will explain transcritical bifurcation in Section ~\ref{sec:Simple zero eigenvalue}. Also, we can see that for parameters $c>3.1998$, and arbitrary $\tau$, $M_{1}$ is always stable.
We should notice that for strong coupling of our FHN neurons the system is excitable, due to saddle-node bifurcation. 
\vspace{-0.2cm}
In the next sections we will study the impact of creation of $M_{1}$ and $M_{2}$, on creation of new stable states. In the next two sections we will first study the bifurcations which are related to equilibrium $M_{2}$, and then bifurcations which are related to equilibrium $M_{1}$. We can see that these bifurcations are necessary for occurrence of multistability. 
\subsection{Bifurcations of $M_{2}$}
\begin{figure}
\centering
\resizebox{0.7\textwidth}{!}{
\includegraphics{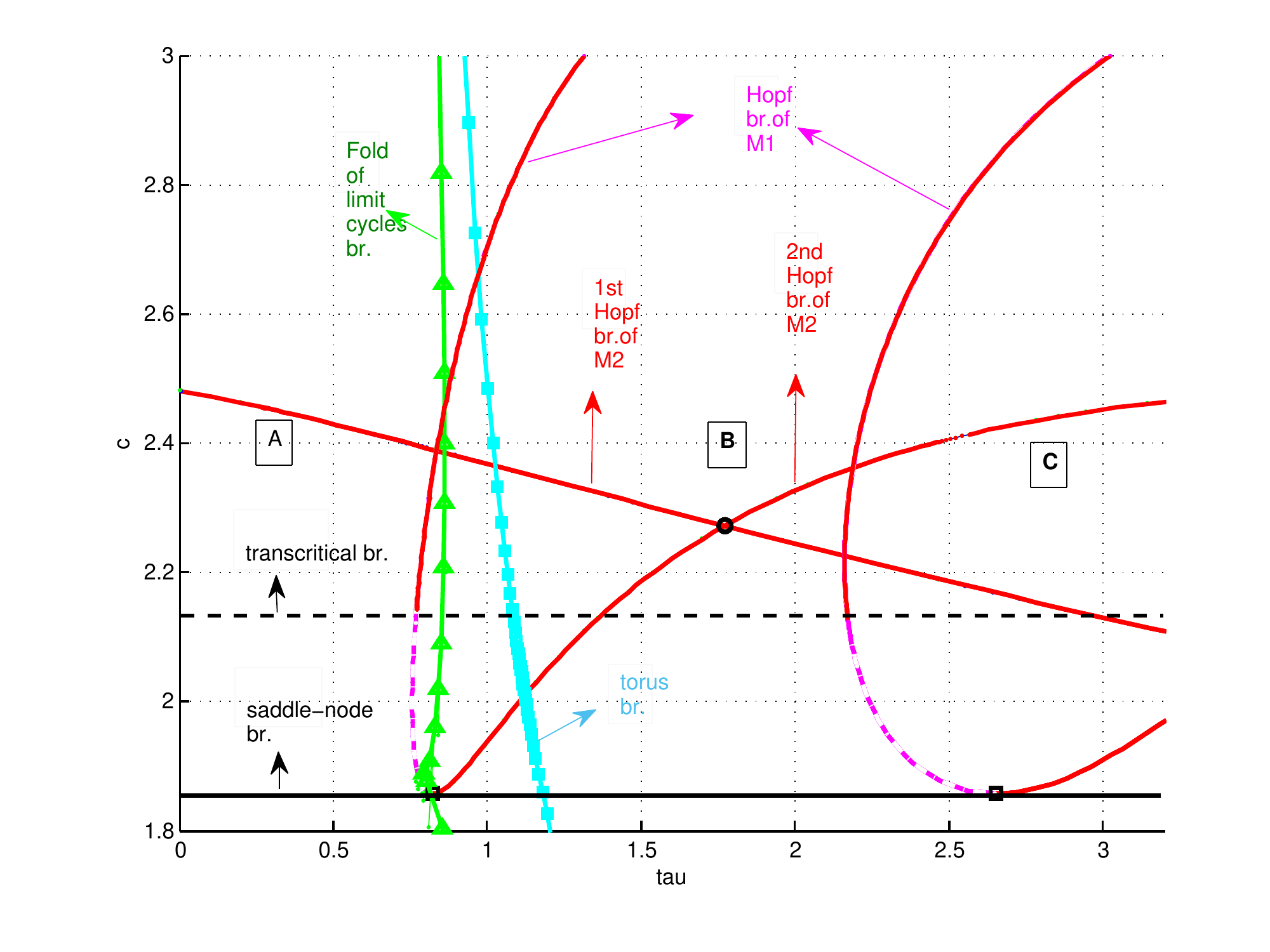}
}
\vspace{-0.7cm}
\caption{\footnotesize Bifurcation diagram of equilibrium $M_{2}$. The solid red lines are sub-critical Hopf bifurcation branches of $M_{2}$ and $M_{1}$. The green line with triangle marker is fold of limit cycles bifurcation branch. The blue line with square marker is torus bifurcation branch. The dash black line is transcritical bifurcation branch. The solid black line is saddle-node bifurcation branch.}
\label{m2}
\end{figure}

\begin{figure}
\centering
\resizebox{0.7\textwidth}{!}{
\includegraphics{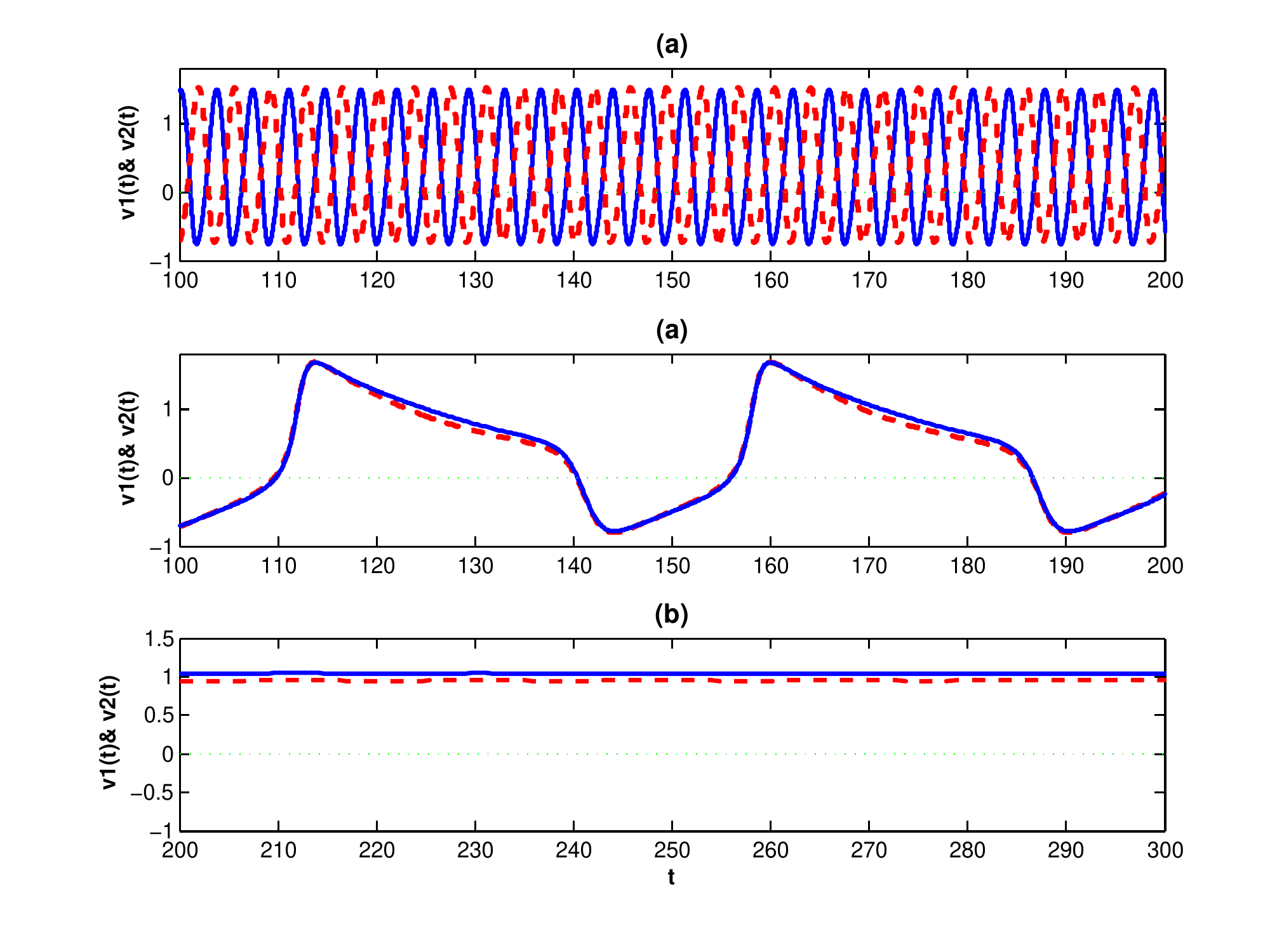}
}
\vspace{-0.7cm}
\caption{\footnotesize Different modes of spiking due to different initial conditions, here $c=2.4$, and $\tau=1.17$. }
\label{3stable}
\end{figure}

\begin{figure}
\centering
\resizebox{1.1\textwidth}{!}{
\includegraphics{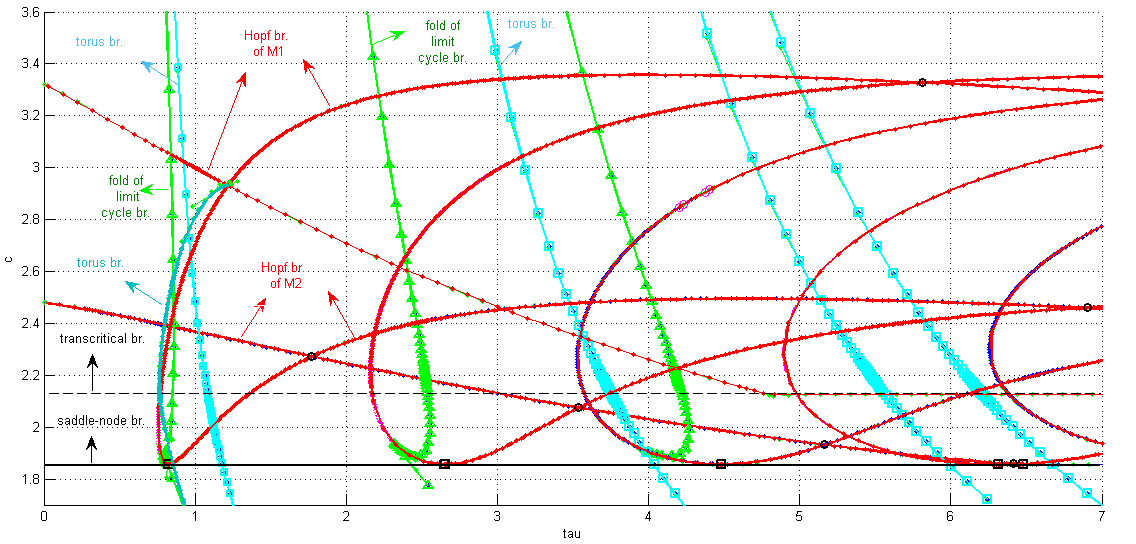}
}
\vspace{-0.7cm}
\caption{\footnotesize Bifurcation diagram of equilibria $M_{2}$ and $M_{1}$. The solid red lines are Hopf bifurcation branches as in Fig. \ref{hopfm1m2}. The green lines with triangle marker are fold of limit cycles bifurcation branches. The blue lines with square marker are torus bifurcation branches. The solid black line is saddle-node bifurcation branch. The dash black line is transcritical bifurcation branch. }
\label{hopfbranchesofm2}
\end{figure}

\begin{figure}
\centering
\resizebox{0.8\textwidth}{!}{
\includegraphics{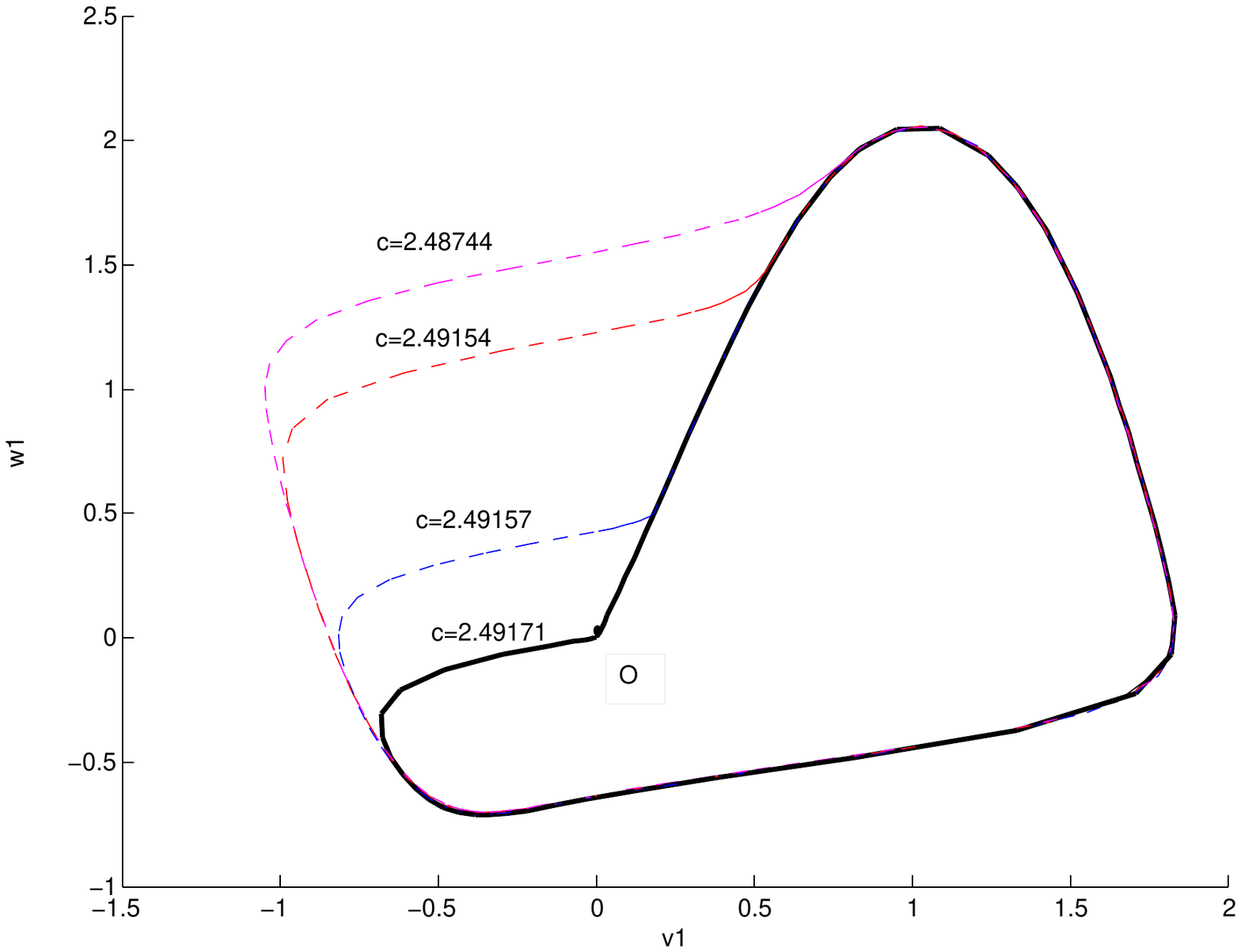}
}
\vspace{-0.9cm}
\caption{\footnotesize The dashed orbits are periodic solutions before creation of big homoclinic orbit. They have different values of $c$, and fixed $\tau$. The solid black orbit is big homoclinic orbit.  }
\label{Bighomoclinic}
\end{figure}

In this section we want to analyze bifurcations of $M_{2}$. As it is shown in the Hopf bifurcations branches of $M_{2}$ in Fig. \ref{hopfm1m2}, the two first branches of Hopf bifurcations of  $M_{2}$ are sub-critical. We will focus on first double-Hopf point of $M_{2}$, and other effective bifurcations around this point, which are depicted in Fig. \ref{m2}, and examine the dynamics in the parameter plane around this point.
 We fix the parameter $c=2.4$ and by changing the parameter $\tau$ we will follow the changes in the dynamics of the system. 
When parameter $\tau$ lies in region A of Fig. \ref{m2}, for example for $\tau=0.4$, $M_{2}$ is unstable and the only stable state of the system is periodic solution, which is correspondent to synchronized activities. By increasing the parameter $\tau$, $M_{2}$ becomes stable by sub-critical Hopf bifurcation. By a little change of parameter $\tau$ fold of limit cycles bifurcation occurs and two limit cycles with index 1 and 2 appear. By increasing the parameter $\tau$ a little more, the limit cycle with index 2 becomes stable through torus bifurcation. This limit cycle is correspondent to anti-phase activities of neurons. Therefore, in region B there are three stable states, $M_{2}$ equilibrium, anti-phase periodic solution, and a big periodic solution which is correspondent to synchronized activities, see Fig. \ref{3stable}. 
When parameter $\tau$ changes from region B to C, $M_{2}$ becomes unstable with sub-critical Hopf bifurcation. Therefore, in region C system is bi-stable with two periodic solutions.\\
By increasing the parameter $\tau$ other branches of fold, Hopf, and torus bifurcations appear. Some of these branches are depicted in Fig. \ref{hopfbranchesofm2}. As a result of the bifurcations listed, other stable limit cycles appear and make the system multistable; similar to what explained in Section ~\ref{sec:multistability}, even for smaller values of parameter $\tau$. \\
As stated in the present section in all the regions A, B, and C there are a big limit cycle which corresponds to synchronized activities. This limit cycle was appeared through Hopf bifurcation of origin. If we continue this periodic solution by varying parameter $c$, we can see that this periodic solution disappears through creation of a big homoclinic orbit, for $c\simeq2.491$, see Fig. \ref{Bighomoclinic}.
\vspace{-0.2cm}
\subsection{Bifurcations of $M_{1}$}
\begin{figure}
\centering
\resizebox{0.9\textwidth}{!}{
\includegraphics{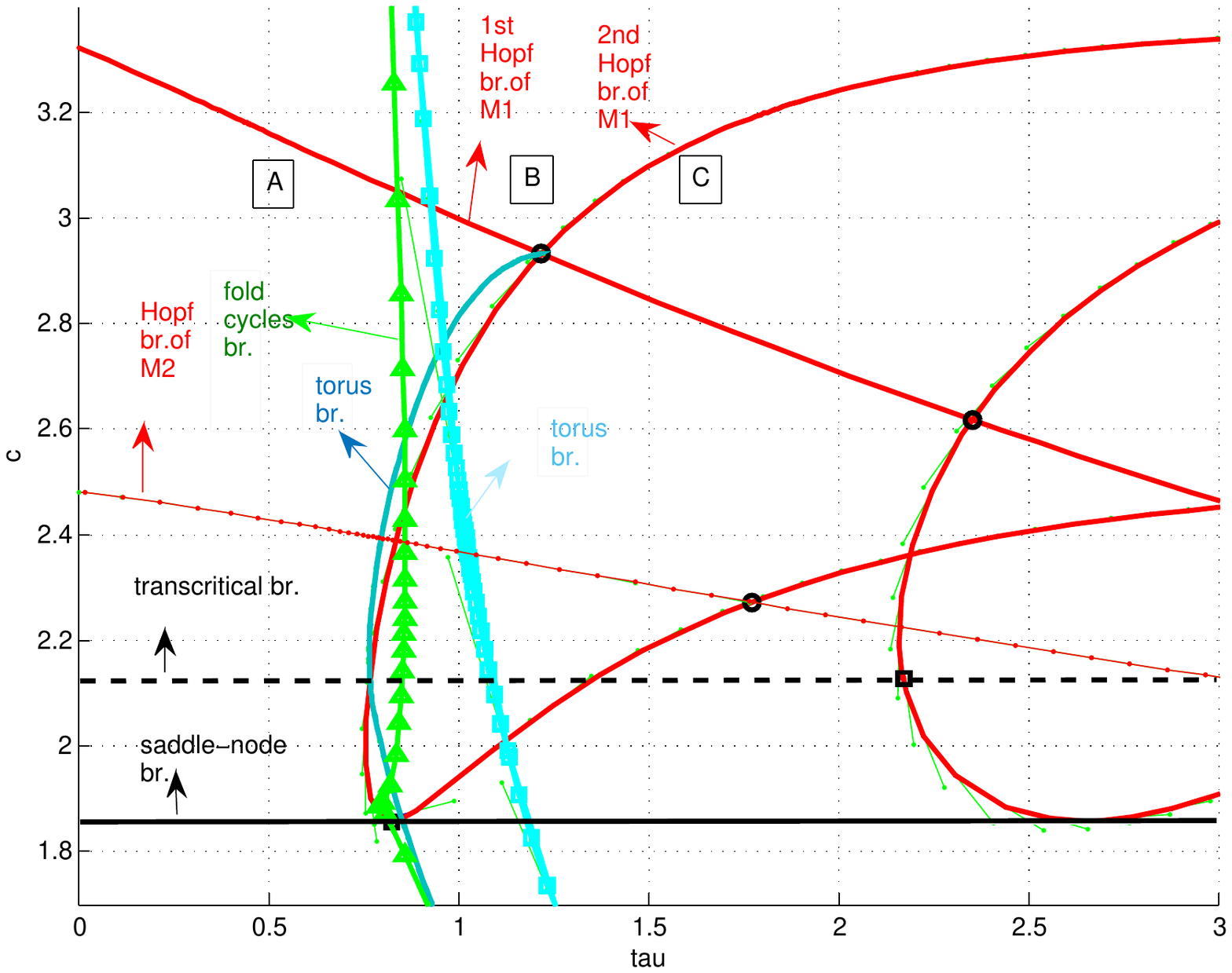}
}
\vspace{-1.1cm}
\caption{\footnotesize Bifurcation diagram of equilibrium $M_{1}$. The solid red lines are Hopf bifurcation branches of $M_{1}$, as in Fig. \ref{hopfm1m2}. The green line with triangle marker is fold of limit cycles bifurcation branch. The blue line with square marker is torus bifurcation branch.  The solid blue line is torus bifurcation branch which emanates from double-Hopf point. The dash black line is transcritical bifurcation branch. The solid black line is saddle-node bifurcation branch. }
\label{m1}
\end{figure}

\begin{figure}
\centering
\resizebox{0.7\textwidth}{!}{
\includegraphics{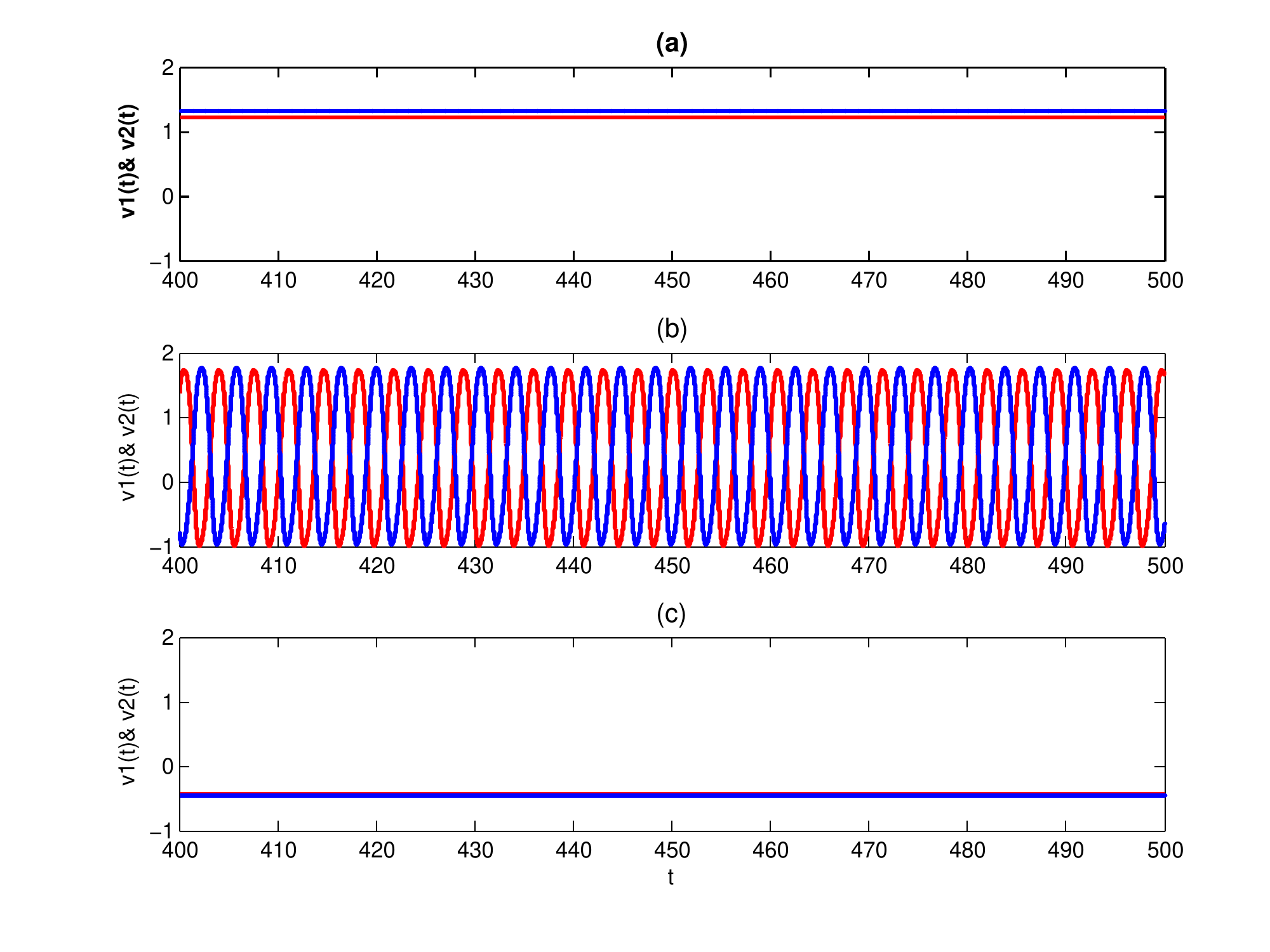}
}
\vspace{-0.4cm}
\caption{\footnotesize Three stable states in region B of Fig. \ref{m1}. Here $c=3.1$, and $\tau=1.25$. a) Eguilibrium $M_{2}$. b)Periodic solution. c) Eguilibrium $M_{1}$ }
\label{regionBofM1}
\end{figure}

In this section we want to analyze bifurcations of $M_{1}$. First we consider the Hopf bifurcations of this point. As it is depicted in Fig. \ref{hopfm1m2}, the Hopf branches of $M_{1}$, change from super-critical to sub-critical by crossing the transcritical bifurcation. We will focus on first double-Hopf point of $M_{1}$, and other effective bifurcations around this point, which are depicted in Fig. \ref{m1}, and examine the dynamics in the parameter plane around this point. The two first branches of Hopf bifurcations of $M_{1}$ around the double-Hopf point are sub-critical. We fix the parameter $c=3.1$, and by changing the parameter $\tau$, we will follow the changes in the dynamic of the system. 
\vspace{-0.2cm}
When parameter $\tau$ lies in region A, for example for $\tau=0.2$, $M_{1}$ is unstable and the only stable state of the system is the equilibrium $M_{2}$. By increasing the parameter $\tau$, $M_{1}$ becomes stable by sub-critical Hopf bifurcation. By a little change of parameter $\tau$ fold of limit cycles bifurcation occurs, and two limit cycles with index 1 and 2 appear. By increasing the parameter $\tau$ a little more, the limit cycle with index 2 becomes stable through torus bifurcation. This limit cycle is correspondent to anti-phase activities of neurons. Therefore, in region B there are three stable states, $M_{1}$ equilibrium, $M_{2}$ equilibrium, and the anti-phase periodic solution, see Fig. \ref{regionBofM1}. 
When parameter $\tau$ changes from region B to C, $M_{1}$ becomes unstable with sub-critical Hopf bifurcation. Therefore, in region C system is bi-stable with a periodic solution and equilibrium $M_{2}$.\\
By increasing the parameter $\tau$ other branches of fold, Hopf, and torus bifurcations appear, Some of these branches are depicted in Fig. \ref{hopfbranchesofm2}. As a result of the bifurcations listed, other stable limit cycles appear and make the system multistable; similar to what explained in Section ~\ref{sec:multistability}, even for smaller values of parameter $\tau$.\\
As stated in Section ~\ref{sec:strong}, the equilibrium $M_{1}$ involves transcritical bifurcation with equilibrium $M_{0}$, we will explain this bifurcation and it's consequences in next section.
\subsection{Transcritical Bifurcation }
\subsubsection{Simple zero eigenvalue}\label{sec:Simple zero eigenvalue}
\begin{figure}
\centering
\resizebox{0.8\textwidth}{!}{
\includegraphics{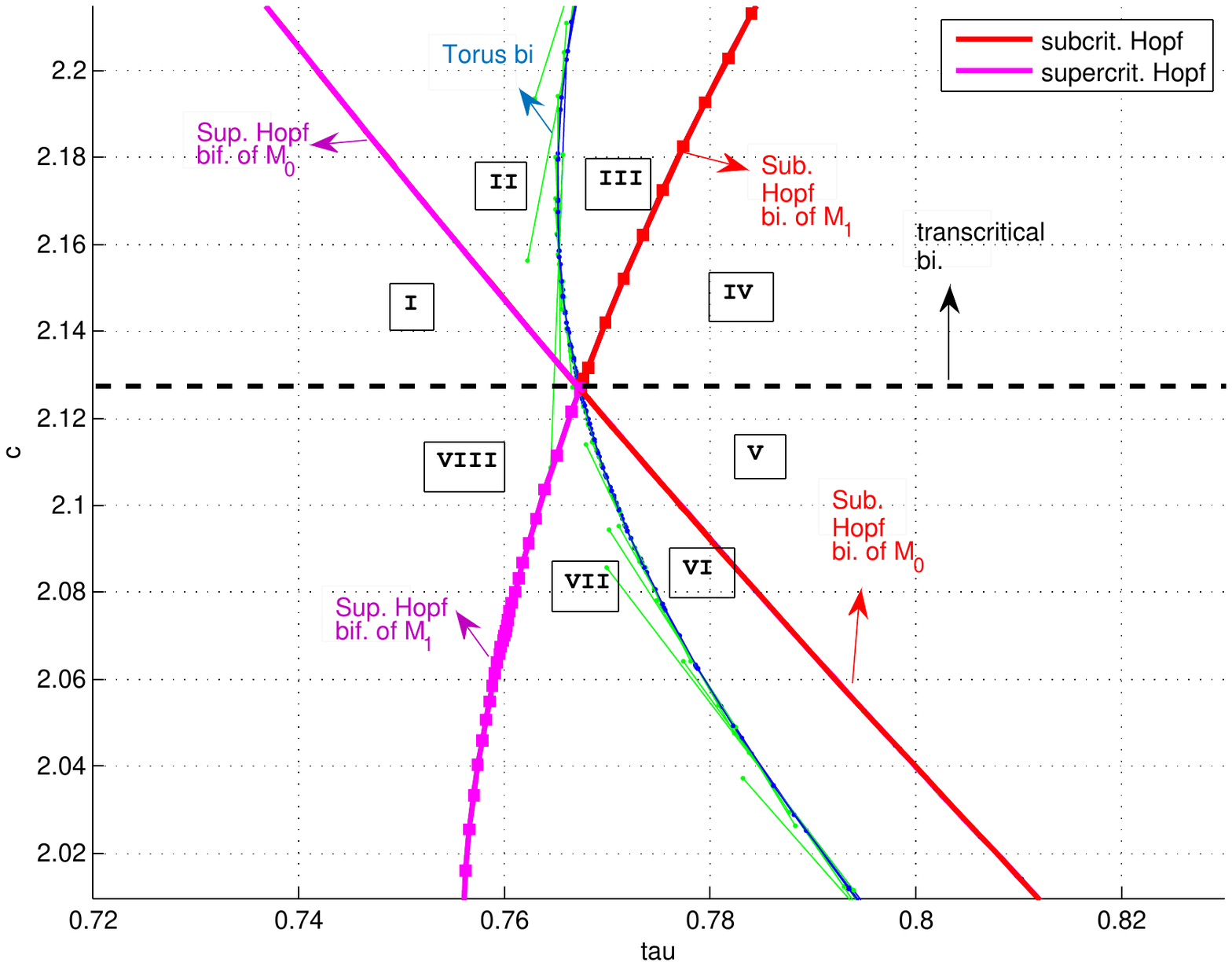}
}
\vspace{-1cm}
\caption{\footnotesize Bifurcation diagram of Hopf-transcritical bifurcation. Magenta lines are super-critical Hopf bifurcations. Red lines are sub-critical Hopf bifurcations. }
\label{transcriticalhopf}
\end{figure}

During study of the system with strong coupling we found that for some values of parameters, by increasing the parameter $c$, the equilibrium $M_{1}$ passes from positive orthant to negative orthant, with exchanging its stability with the origin, but the orthant and stability of $M_{2}$ remains unchanged. We found that it happens through transcritical bifurcation of $M_{0}$. According to Eq. (\ref{2}), and by 
Lemma \ref{transcritical} in Appendix, we can see that for $c\simeq2.12689$ and $\tau\neq 4.7809$, $\lambda=0$ is a simple root of Eq. (\ref{2}). In this situation the trivial equilibrium exhibits a transcritical bifurcation, namely, two equilibria $M_{0}$ and $M_{1}$ collide and exchange stability.	We should emphasize that the equilibrium $M_{2}$ always lies on positive orthant in the considered range of the parameters in this paper. If we consider the impact of parameter $\tau$ and follow the Hopf bifurcations of $M_{0}$ and $M_{1}$, we can see that there are some points in parameter space $(\tau,c)$ in which the Hopf branches of $M_{0}$ and $M_{1}$ meet each other and also the transcritical branch. In this situation Hopf-transcritical bifurcation occurs, one of them is shown in Fig. \ref{transcriticalhopf}. It should be note that the Hopf branch of $M_{1}$ in Fig. \ref{transcriticalhopf}, is the same as the second Hopf bifurcation branch of $M_{1}$ in Fig. \ref{hopfm1m2}, and Fig. \ref{m1}. Also the Hopf branch of $M_{0}$ in Fig. \ref{transcriticalhopf}, is the same as the second Hopf bifurcation branch of $M_{0}$ in Fig. \ref{total}.
 Although all of the involved limit cycles and equilibria around this Hopf-transcritical bifurcation are unstable, they have impact on creation of stable states away from this point. In region I of Fig. \ref{transcriticalhopf}, $M_{0}$ and $M_{1}$ have index 1 and 2 respectively. By passing from region I to II, super-critical Hopf bifurcation occurs for $M_{0}$. Throughout this bifurcation an index-1 limit cycle appears, and changes the index of $M_{0}$ to 3. During passing from region II to III the index-1 limit cycle becomes index-3 by torus bifurcation. It should be note that the stated torus branch in Fig. \ref{transcriticalhopf}, is the same as the torus branch which emanates from double-Hopf point in Fig. \ref{hopfm1m2}. By entering region IV the index-3 limit cycle disappears through sub-critical Hopf bifurcation of $M_{1}$, and makes the index of $M_{1}$ equal 4. Two equilibria $M_{0}$ and $M_{1}$ change their stabilty passing from region IV to V. Therefore in region V, $M_{0}$ and $M_{1}$ have index 4 and 3 respectively. An unstable limit cycle with index 3 appears through sub-critical Hopf bifurcation of $M_{0}$, by entering region VI. By passing from region VI to VII the index of unstable limit cycle changes from 3 to 1 by torus bifurcation. The index-1 limit cycle disappears through super-critical Hopf bifurcation of $M_{1}$. Therefore in region VIII there are two equilibria $M_{0}$ and $M_{1}$ with index 2 and 1.\\ 
We should notice that the first Lyapunov coefficients of both Hopf branches of $M_{0}$ and $M_{1}$ change sign after transcritical bifurcation. Therefore the transcritical branch separates the sub- and super-critical Hopf branches from each other, as is depicted in Fig. \ref{transcriticalhopf}, and Fig. \ref{hopfm1m2}.
\subsubsection{resonance}
\begin{figure}
\centering
\resizebox{0.8\textwidth}{!}{
\includegraphics{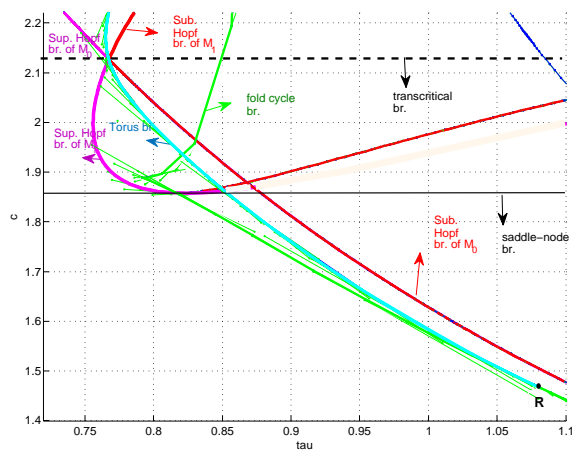}
}
\vspace{-1.5cm}
\caption{\footnotesize The point marked R is $1:1$ resonance point. }
\label{resonance}
\end{figure}

If we decrease parameter $c$, on the first branch of fold of limit cycles bifurcation in Fig. \ref{m1}, and on the torus branch in Fig. \ref{transcriticalhopf},  for $c\simeq1.889$, there is a $1:1$ resonance bifurcation point which is marked R in Fig. \ref{resonance}. In this point the branch of torus bifurcation, which was involved in Hopf-transcritical bifurcation, meets the stated fold cycle branch, \cite{kuznetsov2013elements}. We should notice that the fold of limit cycles branch under this point generates limit cycles with index 2 and 3, while the fold cycle branch above this point generates index-1 and -2 limit cycles. In this point a pair of Floquet multipliers become equal $1$. Actually this $1:1$ resonance is important to link the bifurcations in the case of strong coupling to the bifurcations in the case which coupling strength is smaller.
\subsubsection{Double-zero eigenvalue}
\begin{figure}
\centering
\resizebox{0.7\textwidth}{!}{
\includegraphics{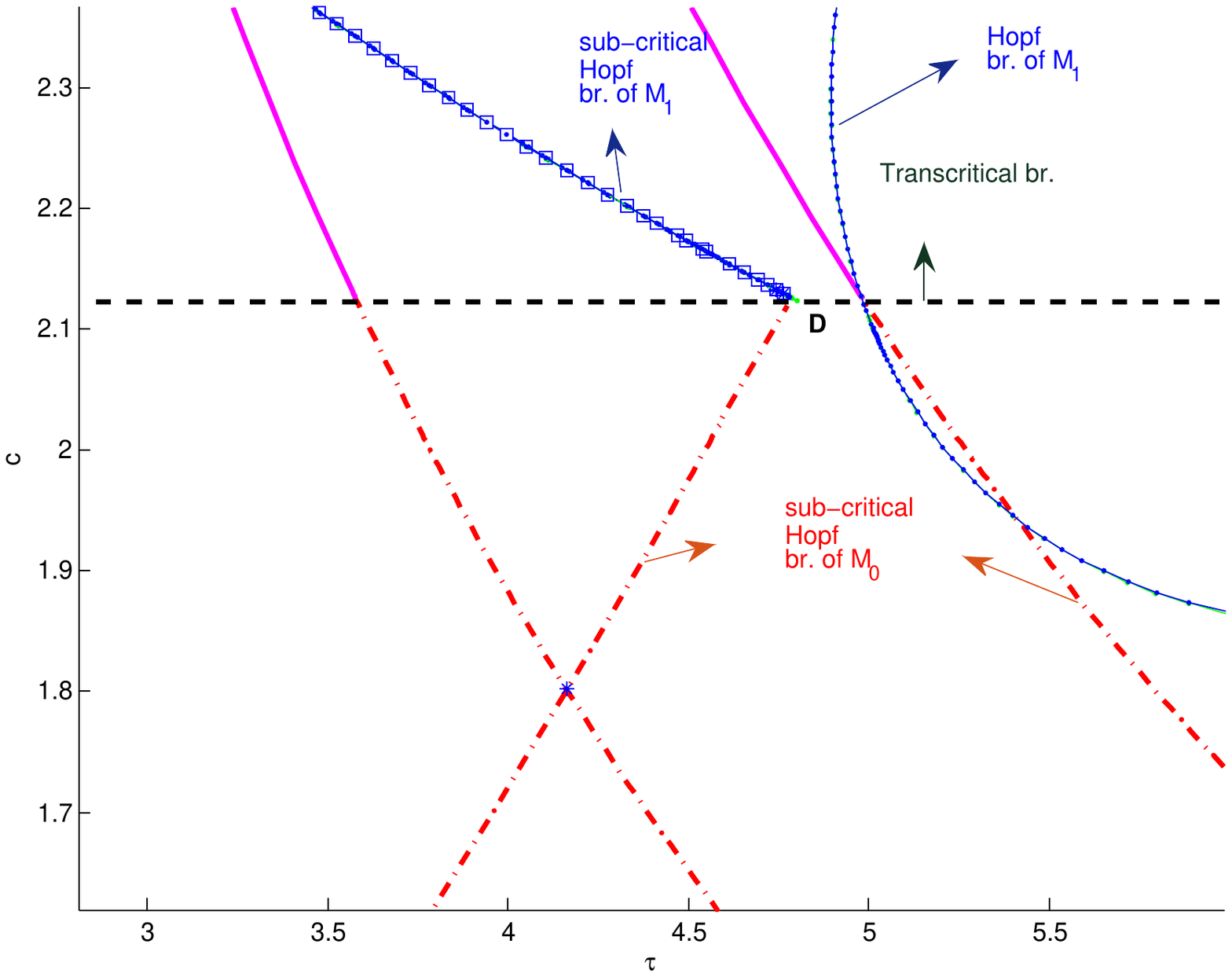}
}
\vspace{-0.9cm}
\caption{\footnotesize The dash-dot red lines are sub-critical Hopf bifurcation branches of $M_{0}$. The blue line with square marker is sub-critical Hopf bifurcation branch of  $M_{1}$. The dash black line is transcritical bifurcation branch. The double-zero bifurcation point is marked D. }
\label{doublezero}
\end{figure}

According to Eq. (\ref{2}), and by Lemma \ref{transcritical} in Appendix, we can see that for $c\simeq2.12689$, and $\tau\simeq 4.7809$, $\lambda=0$ is a double root of Eq. (\ref{2}). Since for $c\simeq2.12689$, and arbitrary $\tau$, $\lambda=0$ is always a solution of Eq. (\ref{2}), transcritical bifurcation occurs for $M_{0}$ and $M_{1}$, as stated in the previous section. Being $\lambda=0$ double root of Eq. (\ref{2}), implies that Hopf bifurcation branches of $M_{0}$ and $M_{1}$, and transcritical branch meet each other in $c\simeq2.12689$, and $\tau\simeq 4.7809$, and Hopf branches end in this point. This phenomenon is called double-zero bifurcation. The corresponding point is marked D in Fig. \ref{doublezero}. It should be note that the Hopf branch of  $M_{1}$ in Fig. \ref{doublezero}, is the same as the first Hopf branch of $M_{1}$ in Fig. \ref{m1}, and Fig. Also the Hopf branch of  $M_{0}$ in Fig. \ref{doublezero}, is the same as the first Hopf branch of $M_{0}$ in Fig. \ref{total}.
\vspace{-0.2cm}
\section{Conclusion}\label{sec:conclusion}
We have used the FitzHugh- Nagumo system which is generic for excitability of type II, i.e., close to a Hopf bifurcation.
We analyzed the dynamics of a system of synaptically coupled FHN neurons with delay, and possible bifurcations of rest points and limit cycles were described.
Besides co-dimension 1 bifurcations such as Hopf,  fold of limit cycles, torus, Chenciner, big homoclinic, saddle-node, and transcritical, we also study co-dimension 2 bifurcations such as double-Hopf, Bautin, fold-torus, Hopf-transcritical, and double-zero, in the parameter plane $(c,\tau)$. Also possible strong resonances of our system are derived. Moreover total bifurcation diagrams for trivial and
non-trivial rest points are reported. We described these bifurcation diagrams by explaining the dynamics in different ranges of parameter plane. Actually this kind of 
bifurcations offer an inevitable key for prediction and detection of complicated sequences of transitions in dynamical systems.  
In this way we have found various dynamical scenarios: quiescence of neurons when trivial rest point is a unigue globally asymptotically stable equilibrium; for weak coupling, different modes of periodic spiking when multistability of periodic orbits or  multistabilty of periodic orbit and torus occurs, possible quiescence and periodic spiking when multistability of non-trivial rest points and periodic solutions happens. Therefore, we showed that with the coupling strength changing, neurons can exhibit rich dynamic and bifurcation behaviors. We also showed that the dynamics can drastically change due to the effect of time delay.
 Missing some patterns of activities is possible, if we only employ numerical simulation. We should notice that these results and detection of sensitive ranges of parameters are interesting from the point of view of applications, since our generic model is representative for a wide range of real-world systems. 
\section{Appendix}
In this appendix we formulate, without proof, some basic Lemmas and theorems about roots of characteristic equation, Eq. (\ref{2}), and the conditions for steady state bifurcations of $M_{0}$.
\begin{lemma}\label{Delta lemma}
Suppose that $ S \geq 0 $, then we have the following results.\\
(i) If $\Delta\geq 0$, then Eq. (\ref{4}) has positive roots if and only if $z_{1}>0$ and $h(z_{1})<0$.\\
(ii) If $\Delta < 0$, then Eq. (\ref{4}) has positive roots if and only if there exists at least one $z^{\ast}\in \{z_{1},z_{2},z_{3}\}$ such that $z^{\ast}>0$ and $h(z^{\ast})\leq 0$.
\end{lemma}
Applying Lemma \ref{Delta lemma}, the Routh-Hurwitz (R-H) criterion, and Ruan and Wei's result \cite{ruan2003zeros} according to Rouche’s theorem, we have the following results.
\begin{lemma}\label{stability}
Assume that $A>0$, $A(B-c^{2})>C-c^{2}(b_{1}+b_{2})$, $D>c^{2} b_{1}b_{2}$, and $[C-c^{2}(b_{1}+b_{2})][A(B-c^{2})-C+c^{2}(b_{1}+b_{2})]>A^2(D-c^{2}b_{1}b_{2})$ are satisfied, ((R-H) hypothesis).\\
(i) If one of the followings holds: (a) $S<0$; (b) $S\geq 0, D\geq 0, z_{1}>0$, and $h(z_{1})\leq 0$; (c) $S\geq0$, $D<0$, and there exists $z^{*}\in \{z_{1},z_{2},z_{3}\}$ such that $z^{*}>0$ and $h(z^{*})\leq 0$, then all roots of (\ref{2}) have negative real parts when $\tau \in [0,\tau_{0})$, such that $\tau_{0}=min \{\tau^{(0)}_{1},\tau^{(0)}_{2},\tau^{(0)}_{3},\tau^{(0)}_{4}\}$.\\
(ii) If the conditions (a)-(c) of (i) are not satisfied, then all roots of (\ref{2}) have negative real parts for all $\tau \geq 0$.
\end{lemma}
Motivated by Lemma 2.3 of the paper of of Ruan and Wei \cite{ruan2003zeros}, Lemma 2.4 of the work of Li and Wei \cite{li2005zeros}, Lemma 2.5 of Hu and Huang \cite{hu2009stability}, and also  Theorem 2.1 of Fan and Hong \cite{fan2010hopf}, we obtain following conclusions.
\begin{lemma}\label{realpart}
Suppose $h'(z_{0})\neq 0$. If $\tau=\tau_{0}$, then $\pm i\omega_{0}$ is a pair of simple purely roots of Eq. (\ref{2}). In addition, $\frac{(d\Re\lambda(\tau))}{d\tau}\vert_{\tau=\tau^{(j)}_{k}}\neq 0$, and the sign of $\frac{(d\Re\lambda(\tau))}{d\tau}\vert_{\tau=\tau^{(j)}_{k}}$ is consistent with that of $h'(z^*_{k})$.
\end{lemma}
Applying Lemmas \ref{stability}-\ref{realpart}, we obtain the following theorem immediately.
\begin{theorem}
Suppose hypothesis (R-H) of Lemma \ref{stability} hold.\\
(i) If non of the conditions (a) $S<0$; (b) $S\geq 0, D\geq 0, z_{1}>0$, and $h(z_{1})\leq 0$; (c) $S\geq0$, $D<0$, and there exists a $z^{*}\in \{z_{1},z_{2},z_{3}\}$ such that $z^{*}>0$ and $h(z^{*})\leq 0$ is satisfied, then the zero solution  of (\ref{1}) is asymptotically stable for all $\tau\geq0$.\\
(ii) If one of the conditions (a), (b), or (c) of (i) is satisfied, then the zero solution of system (\ref{1}), for $\tau \in [0,\tau_{0})$, is asymptotically stable ( $\tau_{0}$ is the parameter defined by the Lemma \ref{stability}).\\
(iii) If one of the conditions (a), (b), and (c) of (i) is satisfied, and  $h'(z^*_{k})\neq 0$, then for $\tau=\tau^{(i)}_{k}, (i=1,2,3,...)$, the system (\ref{1}) undergoes a Hopf bifurcation at $(0,0,0,0)$.
\end{theorem}
Now we want to study the possible steady state bifurcations of the trivial rest point.
Motivated by Lemma 2.1 of the work of Li and Jiang \cite{li2011hopf}, we have the following conclusion on the eigenvalues of  Eq. (\ref{2}).
 \begin{lemma}
 Eq. (\ref{2}) has a zero eigenvalue if and only if $ c^{2}=\frac{(ab_{1}+\gamma)(ab_{2}+\gamma)}{b_{1}b_{2}}$ and $ \tau\neq \frac{b_{1}+b_{2}}{2b_{1}b_{2}}-\frac{a+b_{1}}{2(\gamma+ab_{1})}-\frac{a+b_{2}}{2(\gamma+ab_{2})}$.\\
 $\lambda=0$ is a double root of (\ref{2}), if and only if $ c^{2}=\frac{(ab_{1}+\gamma)(ab_{2}+\gamma)}{b_{1}b_{2}}$, and $ \tau=-\frac{a+b_{1}}{2(\gamma+ab_{1})}-\frac{a+b_{2}}{2(\gamma+ab_{2})}+\frac{b_{1}+b_{2}}{2b_{1}b_{2}} $.
\label{transcritical}
 \end{lemma}


\bibliographystyle{spmpsci}      
\bibliography{ref}   

%

\end{document}